\documentclass[aps,pra,reprint,amsmath,amssymb,superscriptaddress,longbibliography]{revtex4-2} % double spaced for referees
\usepackage{color,graphicx,float,hyperref,stackengine,braket,bm,mathtools,calc,float,eqnarray}
\usepackage{graphicx}% Include figure files
\usepackage{dcolumn}% Align table columns on decimal point
\usepackage{colortbl}
\usepackage{enumitem}% for itemization in roman numerals
\usepackage{placeins}
\newcommand{\oH}{\ensuremath{\mathcal{H}}}

\newcommand{\oA}{\ensuremath{\mathcal{A}}}
\newcommand{\oB}{\ensuremath{\mathcal{B}}}
\newcommand{\oC}{\ensuremath{\mathcal{C}}}
\newcommand{\oD}{\ensuremath{\mathcal{D}}}
\newcommand{\oE}{\ensuremath{\mathcal{E}}}
\newcommand{\oF}{\ensuremath{\mathcal{F}}}

\begin{document}

\title{Wavelet resolved coherence beating in the Overhauser field \\ of a thermal nuclear spin ensemble}
\author{Ekrem Taha \surname{G\"{u}ldeste}}
\author{Ceyhun \surname{Bulutay}}
\email{bulutay@fen.bilkent.edu.tr}
\affiliation{Department of Physics, Bilkent University, Ankara 06800, Turkey}

\date{\today}

\begin{abstract}
This work introduces the so-called synchrosqueezed wavelet transform, to shed light on the dipolar fluctuations of a thermal ensemble of nuclear spins in a diamond crystal structure, hyperfine-coupled to a central spin. The raw time series of the nuclear spin bath coherent dynamics is acquired through the two-point correlation function computed using the cluster correlation expansion method. The dynamics can be conveniently analyzed according to zero-, single-, and double-quantum transitions derived from the dipolar pairwise spin flips. We show that in the early-time behavior when the coherence is preserved in the spin ensemble, the Overhauser field fluctuations are modulated by dipole-dipole-induced small inhomogeneous detunings of nearly resonant transitions within the bath. The resulting beating extending over relatively longer time intervals is featured on the scalograms where both temporal and spectral behaviors of nuclear spin noise are unveiled simultaneously. Moreover, a second kind of beating that affects faster dynamics is readily discernible, originating from the inhomogeneous spread of the hyperfine coupling of each nucleus with the central spin. Additionally, any quadrupolar nuclei within the bath imprint as beating residing in the zero-quantum channel.
The nuclear spin environment can be directionally probed by orienting the hyperfine axis. Thereby, crucial spatial information about the closely separated spin clusters surrounding the central spin are accessible. Thus, a wavelet-based post processing can facilitate the identification of proximal nuclear spins as revealed by their unique beating patterns on the scalograms. Finally, when these features are overwhelmed by either weakly or strongly coupled classical noise sources, we demonstrate the efficacy of thresholding techniques in the wavelet domain in denoising contaminated scalograms.
\end{abstract}

% PhySH
% Research Areas / Quantum information processing
% Physical Systems / Quantum spin models

\maketitle

\section{Introduction}
% Literature Survey
% Central spin and its  applications
A localized electronic spin in a solid-state matrix constitutes an important paradigm as a qubit for quantum information processing purposes \cite{awschalom13,chatterjee21}. Apart from quantum computing, there are numerous experimentally realized applications based on such a spin qubit, such as atomic clock transitions \cite{wolfowicz13}, room temperature quantum simulators \cite{cai13}, time crystals \cite{choi17}, weak measurement protocols \cite{wang19} and quantum sensing of a single nuclear magnon excitation in a quantum dot \cite{jackson21}, to name just a few. In this context the primary concern is the decoherence of the spin qubit, mainly arising from the coupling to background nuclear spins via the hyperfine (hf) interaction \cite{khaetskii02,merkulov02, coish04, cywinskiprl09, ma14, zhang20}. As these nuclear spins interact among themselves through the magnetic dipole-dipole (d-d) interaction, they form an environmental bath for the qubit \cite{hanson08}, and theoretically this corresponds to the central spin (CS) model \cite{gaudin76}. The interaction mechanism between the CS and the nuclear spins due to hf coupling is described via Knight and Overhauser fields: While the former comprises the dynamic magnetic field caused by the CS acting on the nuclear spin environment, the latter is the back-action of the nuclear spins on the spin qubit \cite{urbaszek13}. The d-d spin flips among the spinful nuclei together with the inhomogeneous hf couplings over the nuclear spin bath (NSB) produce Overhauser field fluctuations, and thereby are responsible for the spin diffusion process \cite{klauder62,desousa03, witzel05, witzel06, coish09, cywinski11}.

%Suppresion of nuclear spin noise (DD and DNP)
In quantum information applications, these fluctuations in the Overhauser field are of utmost importance as they decohere the qubit \cite{merkulov02, oleksandr11, sinitsyn12}. A vast number of efforts have been dedicated to the control of the environmental spin noise: among various successful measures, one can mention the dynamical decoupling in terms of sequentially applied pulses \cite{hahn50, viola98, carr54, uhrig07, chekhovich15, cywinski08}, dynamic nuclear polarization through polarization transfer from optically oriented electron spin to nuclear spins to prevent Overhauser fluctuations via hf coupling \cite{imamoglu03, hogele12, jacques09, falk15}, or selectively decoupling and recoupling dipolar spin interactions through a modified version of magic angle spinning \cite{ajoy19}.

%Nuclear spins as quantum registers
As opposed to the unfavorable viewpoint of being the main source of qubit decoherence, nuclear spins in a semiconductor matrix promise to be good candidates for quantum registers because of their much longer coherence times than the electronic spins \cite{taylor03-2, taylor03}. Experimentally, this is realized in quantum dots by the manipulation of collective spin waves \cite{gangloff19}, and enhanced state transfer fidelity rates are reached in the presence of strain \cite{denning19}. In dilute spin systems such as defect centers, this is attained by embedding the CS state to proximal nuclear spins \cite{dutt07,neuman08, fuchs11, taminiau14}. Very recently a 10-qubit spin register with a single-qubit retention time over 75~s and two-qubit entanglement time over 10~s are reported \cite{bradley19}.

% 3D nuclear spin imaging
Another progress that exploits the coherence of nuclear spins in a solid-state matrix strives for extracting a three-dimensional (3D) structural information. For self-assembled quantum dots, this was proposed for gaining insight to their atomistic level composition through optically detected nuclear magnetic resonance (NMR) \cite{bulutay14}. On nanoscale NMR imaging, defect centers have also drawn considerable attention with various sensing and decoupling approaches \cite{ajoy15,kost15,wang16,perunicic16}. The experimental improvement has also been remarkable over a decade, by starting with feature sizes exceeding 10~nm \cite{maze08a,meriles10}, which are further improved by Fourier magnetic imaging techniques \cite{arai15}. Recently, a 3D reconstruction with subangstrom resolution has succeeded with nitrogen vacancy electron spin in diamond \cite{abobeih19}. This methodology depends on the successful isolation of probe spin from environmental nuclear spin noise and recoupling to target spin.

% THEORY: CCE and relatives
Invariably, for all these settings, an in-depth understanding of the nuclear spin fluctuations primarily due to d-d interaction is essential. The theoretical pursuits for this purpose require specialized methods within the so-called pure dephasing model where the electron spin flips are suppressed owing to large energy costs \cite{yao06, cywinskiprl09}. Among the notable choices, there are pair correlation approximation \cite{yao06, liu07, yao07}, cluster expansion \cite{witzel05, witzel06}, linked cluster expansion \cite{saikin07}, cluster correlation expansion (CCE) \cite{yang08, yang09}, disjoint cluster approximation \cite{maze08b, hall14}, and ring diagram approximation \cite{cywinskiprl09, cywinskiprb09}. Out of these, CCE stands out in terms of its easier and wider range of applicability to solid-state spin systems \cite{yang17}. As some latest advancements, CCE is adopted to analyze decoherence of two entangled spin qubits \cite{kwiatkowski18}, and modified to calculate  decoherence of central spins near clock transitions \cite{zhang20}. Furthermore, it is generalized to describe longitudinal relaxation of nitrogen vacancy centers \cite{yang20}, and applied to simulate dynamics of SiC divacancies \cite{onizhuk21, ghosh21}.

%Wavelet Transform
The associated data in either theoretical or experimental quantum coherence studies come in the form of a time series. Thus, introducing additional analysis techniques, beyond the traditional Fourier transform, would be undoubtedly well-founded. Two closely related ones generally applicable to many branches of science and engineering are the short-time Fourier transform and the wavelet transform, where the latter is a multiresolution tool predominantly used in signal and image processing \cite{addison02}. In physics, the wavelet transform received an early welcome, such as in Ising spin chains \cite{phillies95-1, phillies95-2, phillies96}, and was recently promoted by its landmark role in the groundbreaking detection of gravitational waves \cite{abbot16,pham17}. Being an active field, one of the latest developments is that the resolution level in the frequency axis is increased via synchrosqueezed wavelet transform (SST) by reassigning the instantaneous frequencies of the underlying wavelet transform \cite{daubechies11, tary18}.

%This work
In this work we employ the wavelet so-called, scalograms, in particular SST, to shed light on the spectro-temporal characteristics of an unpolarized ensemble of nuclear spins hf-coupled to a CS in a diamond crystal structure \cite{witzel14, ma15}. We keep track of coherence dynamics within the NSB through the two-point correlation function which is computed using the CCE method \cite{ma15}. We show that in the early-time behavior when the coherence is preserved in the spin ensemble, the Overhauser fluctuations are modulated by d-d-induced small inhomogeneous detunings of nearly resonant transitions within the NSB. The resultant beating extending over relatively longer time intervals is readily captured on the scalograms where both temporal and spectral behaviors of nuclear spin noise are unveiled simultaneously. Moreover, a second kind of beating that affects the fast dynamics is also detectable, originating from the inhomogeneous spread of the hf coupling of each nucleus with the CS. The significance of these results is that they provide a crucial spatial information about the spinful nuclei surrounding the CS. In particular, we demonstrate that different realizations in regard to spinful sites have distinct bearings on the scalograms as can be directionally probed by orienting the hf axis. This makes it possible to identify nearest-neighboring spins by means of their distance to the CS, and the alignment of dipolar displacement vector with respect to external magnetic field within the pure-dephasing regime.
Lastly, to render our simplistic model more realistic, we add random telegraph noise to represent classical two-level fluctuators spread over the lattice. Even when the coherence beating becomes totally concealed, its denoising is possible using existing techniques specific to wavelet domain.
Herewith, we assert that wavelet-resolved coherence beating can supply valuable and complementary information that can simplify the task of 3D reconstruction of nuclear spin sites, as recently reported over $^{13}$C nuclei in a nitrogen vacancy center \cite{abobeih19}.

% Plan
The paper is organized as follows. In Sec.~II we present the theoretical information on the interaction Hamiltonian and model details, the CCE method, and the wavelet transform. This is followed in Sec.~III by successive results on how beating patterns emerge and their analysis, including how to combat noise. Section~IV discusses our findings in the light of two recent experiments. Finally, Sec.~V highlights our main conclusions.

\section{Theory}

\subsection{General Hamiltonian}
\begin{figure}[tb!]
		\begin{center}
		\includegraphics[width=1.0\columnwidth]{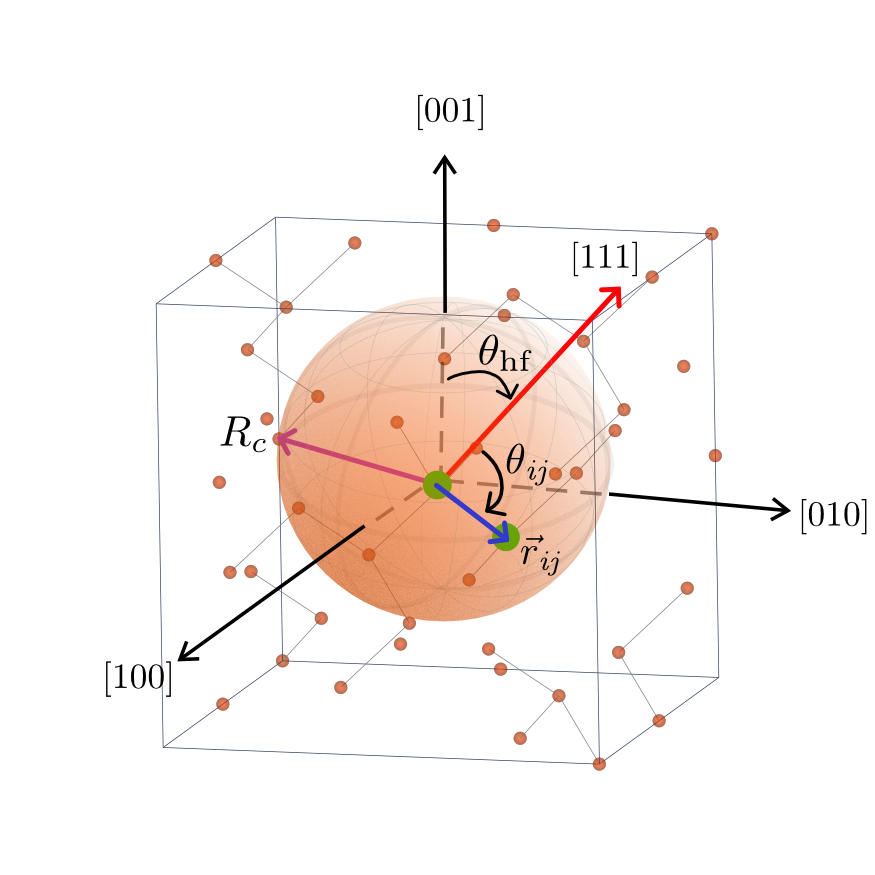}
		\caption{The diamond crystal structure NSB environment. Red and blue  arrows represent the hf axis and the displacement vector, $r_{ij}$, respectively, and the magenta arrow marks the dipole-dipole interaction cutoff radius, $R_c$. $\theta_{\text{hf}}$ designates the angle between the [001] crystallographic direction and the hf axis. The spin quantization axis is taken to be aligned with the hf axis. The spinful nuclei (orange dots) when occuring at nearest-neighbor sites are connected with black lines.}
		\label{crystal}
		\end{center}
\end{figure}

The general Hamiltonian that describes a typical solid-state environment involving an electronic (CS) and background nuclear spins in a quantum dot or a defect center can be expressed as
\begin{equation}
  \oH = w_{\textit{cs}}S^z + \vec{S}\cdot\sum_{j}A_j \vec{I}_j + \sum_j w_j I_j^z + \oH_{\text{d-d}},
\end{equation}
where $\hbar = 1$ and $w_\textit{cs}$ and $w_j$ respectively correspond to Larmor frequencies for CS and the $i$th nuclear spin in the presence of the external magnetic field, known as the Zeeman interactions. $\vec{S}$ ($\vec{I}_j$) represents the central ($j$th nuclear) spin operator. $\oH_{\text{d-d}}$ is the nuclear dipole-dipole interaction to be elaborated below.

The typical frequency (energy) scales of the CS Zeeman interaction under an external magnetic field of 1~T and the hf interaction compare as $10^{11}$~s$^{-1}$ vs $10^6$~s$^{-1}$, as tabulated in Refs.~\cite{witzel08, liu07} for III-V group quantum dots and Si:P donors, so that the electronic Zeeman energy splitting is much larger than the hf coupling to a nuclear spin site, rendering the CS flip process highly suppressed. This enables us to switch to a pure dephasing Hamiltonian by dropping the nonsecular part of the hf interaction \cite{giedke06}, and it becomes convenient to choose the spin quantization axis parallel to the magnetic field vector. Thus, in this work the spin quantization axis (which identifies $I^z$) is taken to be aligned with the hf axis (see Fig.~\ref{crystal}), where the latter will be oriented in various directions to study its consequences. Additionally, in the strong Zeeman regime the secular part of $\oH_{\text{d-d}}$ can also be dropped by means of adiabatic approximation \cite{slichter90, abragam99}, as these terms become fast oscillating, which is going to be discussed separately in Sec. \ref{highBfieldregime}. However, for the general part of our analysis we keep non-secular parts of the d-d interaction to investigate NSB dynamics under weak magnetic field (say, around $10$~mT), where the pure dephasing Hamiltonian can still be valid \cite{cywinskiprl09,cywinskiprb09}.

The Hamiltonian conditioned on the CS state is written as \cite{ma15}
\begin{equation} \label{Genel Ham}
\oH^{(\pm)}= P_{\pm} \beta^{z}+\oH_{\text{d-d}},
\end{equation}
where $\beta^z = \sum_i \beta_i^z = \sum_i A_i I_i^z$  is the bath operator, $A_i$ and $I_i^z$ are the hf interaction coupling constant and nuclear spin operator's component along the quantization axis for the $i$th lattice site, respectively, and $P_{\pm}=\left\langle\pm\left|S^{z}\right| \pm\right\rangle$, with $ |\pm\rangle$ representing CS eigenstates along the hf axis. It is important to note that in Eq.~(\ref{Genel Ham}), we neglect the nuclear Zeeman interaction term, since its contribution to the NSB dynamics is insignificant in the low-magnetic-field regime. However, the direction of the magnetic field remains to be critical as it determines the hf axis. So, the d-d Hamiltonian in SI units the under low-field-regime reads
\begin{equation}\label{alp_closed}
  \oH_{\text{dd}} = \sum _{i>j} \dfrac{\mu _0}{4 \pi} \dfrac{\hbar \gamma_i \gamma_j }{|\vec{r}_{ij}|^3} ( \oA + \oB + \oC + \oD + \oE + \oF ) ,
\end{equation}
where $\mu_0$ is the permeability of free space, $\gamma_i$ and $\gamma_j$ are gyromagnetic ratios, and $\vec{r}_{ij}$  is the vector that connects two nuclear spin sites. $\oA,\oB,\oC,\oD,\oE$ and $\oF$ operators constitute the so-called dipolar alphabet \cite{slichter90, abragam99} having the explicit forms
\begin{subequations}
\label{alphabet}
 \begin{eqnarray}
 \oA &=& I_{i}^z I_{j}^z (3 \cos^2 \theta_{ij} -1),\\
 \oB &=& (I_i^+I_j^- + I_i^-I_j^+ )\dfrac{1-3 \cos^2 \theta_{ij}}{4}, \\
 \oC &=& (I_i^+I_j^z + I_i^zI_j^+ )\Big ( \dfrac{3}{4}   \sin2\theta_{ij} \Big
)e^{-i\phi_{ij}},\\
 \oD &=& (I_i^-I_j^z + I_i^zI_j^- )\Big ( \dfrac{3}{4}   \sin2\theta_{ij} \Big
)e^{+i\phi_{ij}}, \\
 \oE &=& I_i^+I_j^+\Big ( \dfrac{3}{4}   \sin^2\theta_{ij} \Big )e^{-2i\phi_{ij}},\\
 \oF &=& I_i^-I_j^-\Big ( \dfrac{3}{4}   \sin^2\theta_{ij} \Big )e^{+2i\phi_{ij}}.
\end{eqnarray}
\end{subequations}
where $I^{\pm}$ are spin creation and annihilation operators, and $\theta_{ij}$ ($\phi_{ij}$)  is the polar (azimuthal) angle defined by the quantization axis and $\vec{r}_{ij}$; see Fig.~\ref{crystal}.

Following Ref.~\cite{ma15}, nuclear spin fluctuations coupled to the two-level spin system can be expressed by a two-point correlation function in the Heisenberg picture as
\begin{equation}
  C(t) = \langle \beta^z(t)\beta^z(0) \rangle = \langle e^{i \oH_e t} \beta^z (0)e^{-i \oH_e t}\beta^z(0) \rangle,
\end{equation}
where $\oH_e$ is the effective Hamiltonian governing the time evolution. According to the prescription in Ref. \cite{ma15}, averaging out the $|\pm\rangle$ hf fields so that,
\begin{equation}\label{effective_Hamiltonian}
  \oH_e = \dfrac{|P_+| + |P_-|}{2}\beta^z +  \oH_{\text{d-d}} ,
\end{equation}
leads to the best agreement with experiments.

\subsection{Other Model Details}
The nuclei presiding under the envelope of a confined electron's wave function in a typical solid-state matrix experience large variations in hf coupling depending on the spatial position of the relevant nuclear spin site \cite{schliemann03, coish04}. This spread within the NSB, as quantified by the standard deviation, $\sigma_{\text{hf}}$, crucially affects the decoherence of CS \cite{Guldeste18}. In our model, the hf coupling constants for bath spins are distributed as
\begin{equation} \label{hyperfine}
  A_i = A_0 e^{-r_i^2/L_0^2},
\end{equation}
here, $A_0$ is hf coupling strength at the peak magnitude location of the electron wave function (taken as the origin of the crystal), $r_i$ represents the distance of the $i$th nuclear spin site from the origin, and $L_0$ is the electron confinement radius, chosen for the purposes of a model study as $3.4$~nm. Its value will be critical in a real application. $A_0$ in Eq.~(\ref{hyperfine}), which determines the hf energy scale, is set to $10^4~E_{dd}$, where $E_{dd}$ corresponds to the d-d energy of two spins separated by a bond length,
\begin{equation}
  E_{dd} =\dfrac{\mu _0}{4 \pi} \dfrac{\hbar \gamma_i \gamma_j }{(a_0\sqrt{3}/4)^3},
\end{equation}
with $a_0$ being the lattice constant of the diamond crystal structure. Two different lattice constants (of silicon and diamond) are considered in showcasing our results. In a computational box of dimensions, say $10~a_0\times 10~a_0\times 7~a_0$, there are $5\/600$ nuclear sites available. Typical defect center concentrations are in the few ppm level \cite{giri19}; therefore only one CS can safely be assumed in such a computational volume. We work in the few percent spinful abundance ratio $\rho$, which is typically the case in group-IV semiconductors, such as in natural silicon where the spin-1/2 isotope $^{29}$Si has abundance $\rho\approx 0.05$ \cite{coish04}, or nitrogen vacancy centers in diamond where $^{13}$C abundance is $\rho \approx 0.01$ \cite{fuchs11}.

\subsection{Cluster Correlation Expansion}
CCE is a many-body technique that is well suited to study NSB fluctuations \cite{yang08, yang09, ma15}.
Even though the procedure is detailed in Refs.~\cite{ma15,yang17}, for the sake of completeness it is beneficial to address CCE method in the nuclear spin noise context briefly. This noise can be probed by means of a two-point correlation function
\begin{equation}
  C(t) = \sum_{ \{i \}} \tilde{C}_{\{i \} }(t) + \sum_{ \{i,j \}} \tilde{C}_{\{i,j \} } (t) + \dots,
\end{equation}
where $\tilde{C}_{ \{ \zeta \} } (t)$ represents the contribution from cluster $\{i,j\}$, so that the effects of all possible irreducible subclusters are excluded as
\begin{equation}
\tilde{C}_{\zeta}(t) = C_{\zeta}(t) - \sum_{\zeta' \subset \zeta} \tilde{C}_{\zeta'}(t).
\end{equation}

By carrying out cluster correlation expansion up to a certain cluster size $M$ (i.e., CCE-M) convergent results can be obtained. The time evolution of the cluster $\zeta$ can be calculated via $C_{\zeta}(t) = \langle e^{i \oH_e^{\zeta} t} \beta^z (0)e^{-i \oH_e^{\zeta} t}\beta^z(0) \rangle$, where $ \oH_e^{\zeta} $ includes interactions only for the relevant cluster. We employ a thermal ensemble at infinite temperature for the NSB (\textit{i.e.} $\rho_B \propto \otimes_{i=1}^{i=N} \mathcal{I}_i $) so as to have vanishing cross-correlation terms in a spin autocorrelation function as calculated using the CCE method \cite{ma15}.
\begin{figure*}[t!]
		\begin{center}
		\includegraphics[width=1.5\columnwidth]{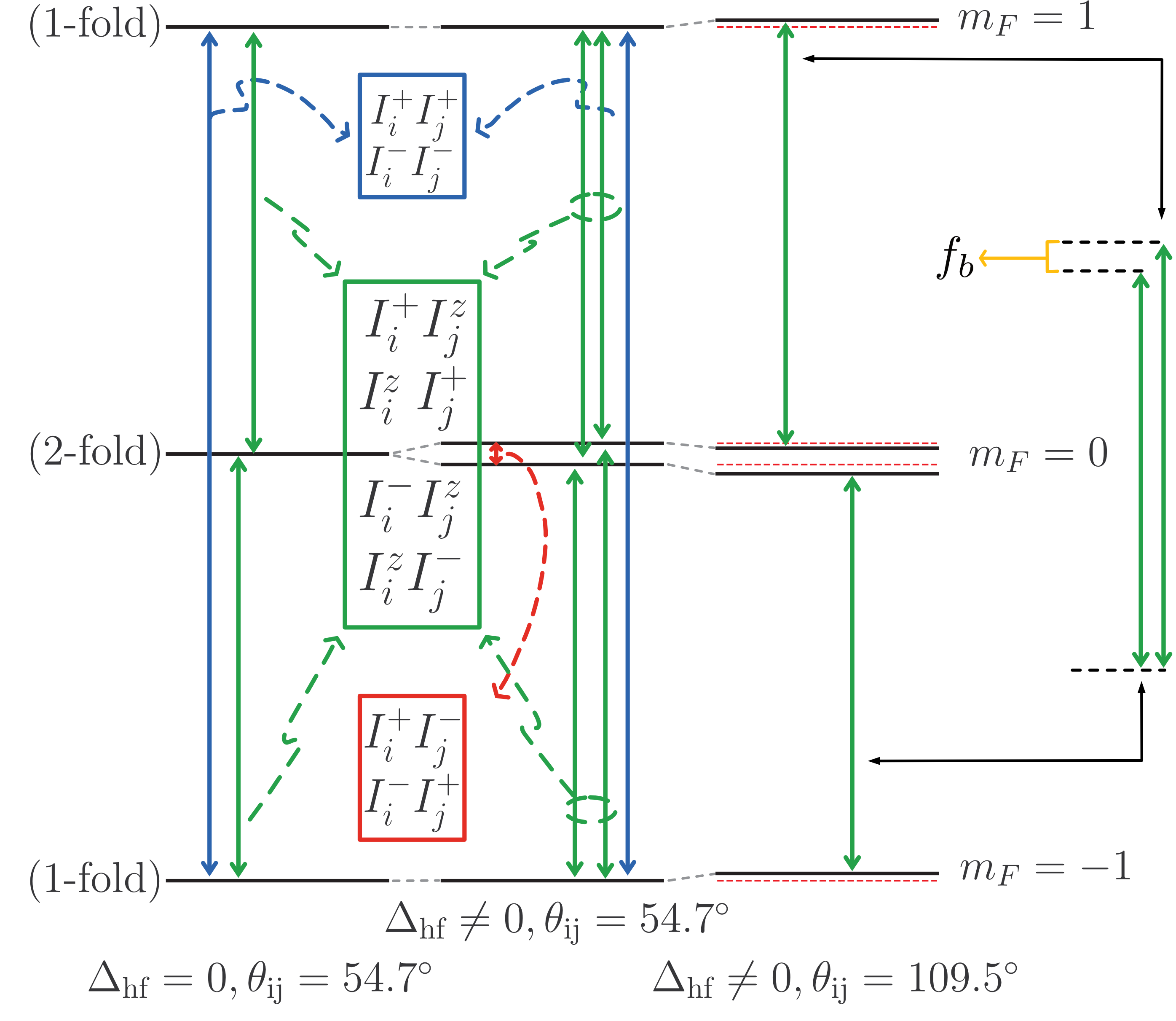}
		\caption{Energy level diagram for typical two nearest-neighboring spin-1/2 nuclei for various $\Delta_{\text{hf}}$ and $\theta_\text{ij}$ values. $f_b$ represents the frequency detuning of two single-quantum transitions for $\Delta_{\text{hf}}\neq0$ and $\theta_{\text{ij}}=109.5^\circ$ case, in which energy levels are slightly shifted due to nonvanishing dipolar alphabet term $\oA$.}
		\label{splitting}
		\end{center}
	\end{figure*}

\subsection{Multi-Resolutional Analysis}
\subsubsection{Continuous Wavelet Transform}
In providing multiresolution analysis, both short-time Fourier transform and continuous wavelet transform (CWT) rely on windowed kernels to localize the time and frequency component of a given signal, simultaneously. A key advantage of CWT over the former is that it inherently utilizes a scalable window depending on the frequency components of interest instead of using fixed windows \cite{addison02}. The dimensions of the window employed depend on the frequency component of interest, so that the fast oscillatory components have low resolution on the frequency axis and high resolution on the time axis, and vice versa.

For a  given signal, $x(t)$, CWT employs a basis function which is localized both in time and frequency domain. There are various basis functions, $\varphi(t)$, or so-called wavelets, available in the literature such as the Mexican hat, bump, Morlet, etc., provided that they have finite energy and zero mean \cite{addison02, stephane09}. In this work we utilize the \textit{bump} wavelet which yields superior localization in the frequency domain with respect to time domain. It is expressed in the Fourier domain as \cite{sylvain12}
\begin{equation}
 \hat{\varphi}(aw) = \exp \left( 1-\dfrac{1}{1-(aw-\mu)^2/\sigma^2} \right)\chi_{\left[(\mu-\sigma)/a, (\mu+\sigma)/a\right]},
\end{equation}
Here, $\hat{\varphi}$ denotes the Fourier transform of the $\varphi$, $\sigma$ and $\mu$ determine the width of the window of time-frequency localization and peak frequency of the wavelet, respectively, and $\chi_{\left[(\mu-\sigma)/a, (\mu+\sigma)/a\right]}$ is the indicator function for the interval $(\mu-\sigma)/a \leq w \leq (\mu+\sigma)/a$.

The wavelet transform  of a signal, $x(t)$ is given by
\begin{equation}
  W_x(a,b) = \dfrac{1}{\sqrt{a}} \int_{-\infty}^{+\infty}x(t)\,\varphi^* \left( \dfrac{t-b}{a} \right)dt,
\end{equation}
here $a$ is the contraction (or dilation) scale parameter and it is inversely proportional to frequency, whereas $b$ provides translation over the time axis of the wavelet. In our work, we plot the scalograms choosing the vertical axis as frequency rather than the scale, for convenience.

\subsubsection{Synchrosqueezed Wavelet Transform}
It is possible to increase the frequency axis resolution of the CWT by means of reallocation of instantaneous frequencies as \cite{daubechies11, tary18}

\begin{equation} \label{sswt1}
  w_x(a,b)=-i(W_x(a,b))^{-1} \dfrac{\partial}{\partial b} W_x(a,b).
\end{equation}
Then, the SST can be written as
\begin{equation}\label{sswt2}
  T_x(w,b) = \int_{A(b)} W_x(a,b)a^{-3/2} \delta(w(a,b)-w)da,
\end{equation}
here, $A(b) = \{a; W_x(a,b)\neq0 \}$ implying that integral is carried out for each $a$ provided that the wavelet transform has a nonvanishing component. As long as the condition is satisfied for frequency separation components of the signal $x(t)$, the CWT can be sharpened remarkably in the frequency axis while the time resolution remains unchanged. We should note that since the bump wavelet (and hence $W(a,b)$ and $T_x(w,b)$) has imaginary parts in the time domain, we will only display its modulus.

\section{Results}

\subsection{Dipole-dipole transitions in a two-spin cluster}

In this subsection we explain the significance of each parameter for the case of a two-spin cluster. The displacement vector $\vec{r}_{ij}$ in the d-d Hamiltonian determines the strength of each alphabet term. The polar angle $\theta_{ij}$ which appears in each of the alphabet terms in Eq.~(\ref{alphabet}) is defined as between $\vec{r}_{ij}$ and the hf axis taken along the $z$ axis which we also choose as our quantization axis for spin basis representation as indicated in Fig.~\ref{crystal}. For a two-spin cluster, when the detuning in the hf interaction is set to zero ($\Delta_{\text{hf}}=0$), the energy level diagrams become completely degenerate if $\vec{r}_{ij}$ is elongated in the $[111]$ direction.

Next, we proceed to how each d-d alphabet term in Eq.~(\ref{alphabet}) is operational on the level transitions among the two-nuclear spin states $|m_F\rangle$ as depicted in Fig.~\ref{splitting}. The energy level splittings in the presence of hf and d-d interaction are decisive in the resultant NSB dynamics. The mean value of hf interaction couplings and the $\oA$ term in the d-d alphabet determine the energy level splittings as the diagonal entries of the $\oH_{\text{bath}}$ matrix.
Uneven hf couplings lift the degeneracy of $|m_F = 0 \rangle$ doublet linearly and enable the intraband zero-quantum (0Q, $\Delta m_F =0$) transitions induced by spin-conserving term $\oB$ of the d-d alphabet as long as it is not masked by the polar angle $\theta_{ij}$ term. The single-quantum (1Q, $\Delta m_F = 1$) transitions are triggered by $\oC$ and $\oD$ which give rise to two distinct frequencies in the $\Delta_{\text{hf}}\neq0$ regime. Lastly, $\oE$ and $\oF$ are responsible for the double-quantum (2Q, $\Delta m_F = 2$) transitions so that they exhibit the same dynamics whether the $\Delta_{\text{hf}}$ vanishes or not.
\begin{figure*}[t!]
    \begin{center}
      \includegraphics[width=1.7\columnwidth]{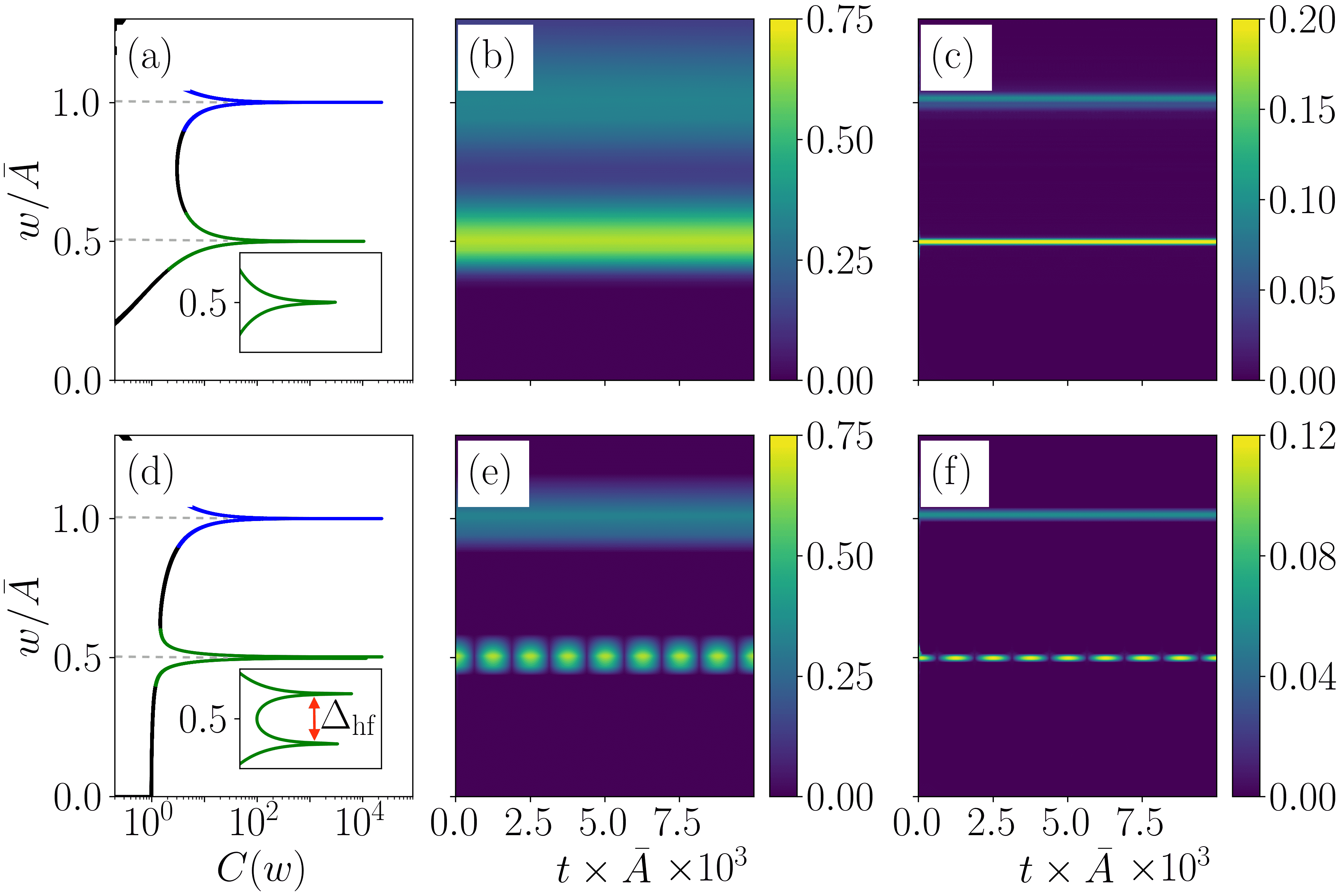}
      \caption{Frequency and time-frequency analysis for two spin-1/2 nuclear spins for $\theta_{\text{hf}} = 0^\circ$. The upper and lower panels display the two-point correlation function behavior for $\Delta_{\text{hf}} = 0$ and $\Delta_{\text{hf}} = 0.01 \bar{A}$, respectively. (a) and (d) show the power spectra of $\bar{C}(t)$, (b) and (e) display their corresponding continuous wavelet scalograms, and (c) and (f) illustrate their SSTs. }
      \label{2spinbath}
    \end{center}
\end{figure*}
\subsection{Time-frequency analysis}
The two-point correlation function, $C(t)$, can be interpreted as the collective contributions of various $\Delta m_F$ transitions that occur in small-spin clusters. For this purpose, the so-called bump wavelet transform of $C(t)$ is a helpful tool to uncover the NSB dynamics as it renders information in both temporal and spectral behavior of nuclear spin noise simultaneously \cite{addison02, stephane09}. But, prior to the wavelet transform,  $C(t)$ is normalized as

\begin{equation} \label{norm}
  \bar{C}(t) = \dfrac{C(t) - \langle C(t) \rangle_t}{C(0)- \langle C(t) \rangle_t},
\end{equation}
where $\langle \dots \rangle_t$ represents the time average so that $\bar{C}(t=0) = 1$.
For convenience, throughout our work normalized time $\bar{t}=t \bar{A}$, and (angular) frequency $\bar{\omega}=\omega/ \bar{A}$ are used, where $\bar{A} = \sum_i A_i/N$ is the mean value of hf couplings for $N$ nuclear spins.

For an $N=2$ spin-1/2 cluster, Fig.~\ref{2spinbath} displays the corresponding frequency spectra (first column), bump wavelet scalogram (second column), and the SST (third column). Note the same colors being used for the relevant portions of the power spectra, as the transition arrows in Fig.~\ref{splitting}. According to our normalized frequency scheme, 1Q transitions occur at $\bar{\omega} = 0.5$, while the 2Q transitions are centered at $\bar{\omega}=1.0$. The spin flip-flop terms are not operative in these scalograms since we choose a configuration from the diamond structure's nearest-neighboring spins having the $\theta_{ij} = 54.7^\circ$. 0Q transitions (located around $\bar{\omega}=0$) arise from dipoles aligned along, say the [001] direction, as will be demonstrated in the following figures. Most notably, under non-zero detuning ($\Delta_{\text{hf}}\neq 0$) a beating pattern appears in the 1Q channel of the wavelet transform as seen in Fig.~\ref{2spinbath} (e). The probability amplitude interference of two distinct transitions separated by an amount of $\Delta_{\text{hf}}$ as indicated in the inset directly determines the beating note in the CWT. Resolving this beating pattern due to $\Delta_{\text{hf}}$ of two nearest-neighboring spins can be hindered for long-time simulations, and appears as a horizontal stripe (see, for example, Fig.~\ref{cce_comp}).

Another observation from Fig.~\ref{2spinbath} (b) and (e) is that frequency localization of the CWT intrinsically becomes weaker as the frequency of interest gets wider. Even the 1Q transition channel, which has relatively good localization in frequency when compared to 2Q transitions, gets somewhat blurred. To better resolve nearby frequency components in large NSBs, we switch to SST as illustrated in Fig.~\ref{2spinbath} (c) and (f). In the remainder of this work, we employ SST by means of Eqs.~(\ref{sswt1}) and (\ref{sswt2}) in which the resolution in frequency axis is highly improved, without hampering the time localization.

\subsection{Convergence tests}
\begin{figure}[tb!]
     \begin{center}
       \includegraphics[width=1\columnwidth]{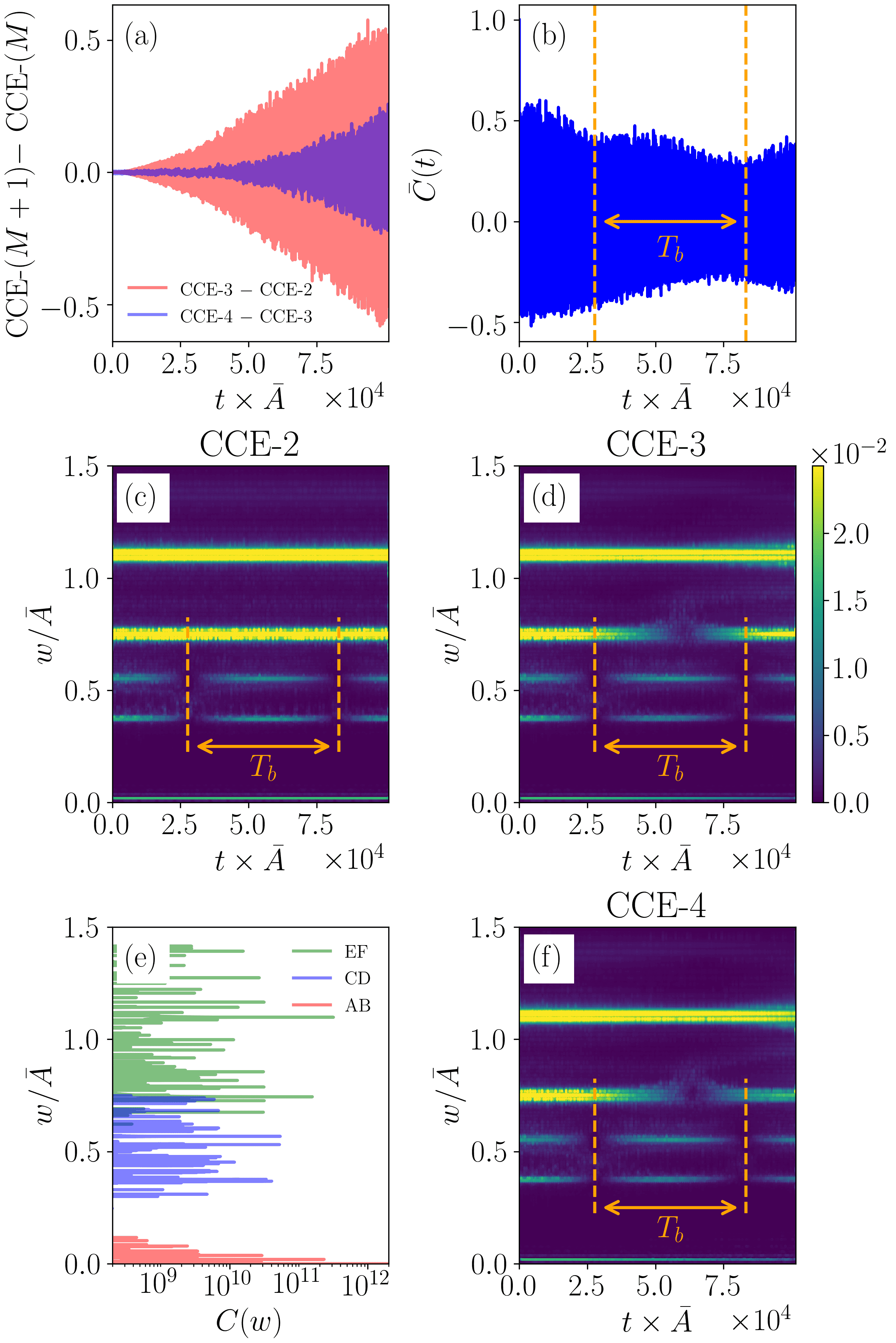}
       \caption{Nuclear spin noise fluctuations under different CCE orders for the spin-$1/2$ bath with $\bar{A} \approx 13.5$~MHz, $\sigma_{\text{hf}} \approx 0.25 \bar{A}$, $\theta_{\text{hf}}  \approx 54.7^\circ$, $N=112$, $\rho= 0.02$,  $a_0 = 5.43$~\r{A} (silicon), and $R_c = 2.7a_0$. (a) Deviation from higher-order CCE of $\bar{C}(t)$; (b) temporal behavior of normalized two-point correlation function, $\bar{C}(t)$, calculated via CCE-4. (c), (d), and (f) display the SST scalograms of spin fluctuations for different CCE orders from CCE-2 to CCE-4. (e) Spectral behavior of $\bar{C}(w)$  calculated with CCE-4.}
       \label{cce_comp}
     \end{center}
 \end{figure}
The CCE, as a many-body technique, requires truncating the expansion up to a certain cluster size \cite{yang08, yang09}; hence we need to assure that convergent results are produced. In Fig.~\ref{cce_comp} (a), we display the convergence of different CCE orders for a spin-1/2 bath up to clusters of four nuclear spins, since the contributions from clusters of size five and six become rather small within the typical baths that we consider \cite{witzel12}. The higher-order correlations that would occur in NSB are especially suppressed, thanks to the low spinful abundance ratio $\rho = 0.02$ which spatially isolates nuclear spin clusters. The NSB short-time dynamics is well described by the pair correlations, i.e., CCE-2. However, for long-time dynamics, as observed from Fig.~\ref{cce_comp} (a), CCE-2 starts to deviate from higher orders. In Fig.~\ref{cce_comp} (c), (d) and (f) we plot different CCE orders; at a first glance, there are two beatings with period $T_b = 1/ f_b$ in 1Q transitions due to two different nearest-neighboring spin clusters. This can be explained through the small detunings in energy levels because of the secular $\oA$ term in the d-d alphabet, as marked with $f_b$ in Fig~\ref{splitting}.

There is also another beating around $w/\bar{A}=0.75$ which is registered by both CCE-3 and CCE-4, while CCE-2 is unable to capture it. This beating profile belongs to 2Q transitions and is induced by the multiple spins within a close range such as clusters of three or more spins. Similarly to the  first-nearest-neighboring spins, three or more spins which are in proximity can constitute multiple 2Q transitions that are slightly detuned from one another and form a beating pattern. Additionally, the structure of the beating profile directly depends on the physical orientation of multiple spins in the host matrix as well as the hf angle.
If we compare the oscillatory profile of the spin noise envelope in Fig~\ref{cce_comp} (a) and beating patterns of (f) which are both calculated with CCE-4, it can be clearly seen that the envelope of $\bar{C}(t)$ follows the beat patterns of two 1Q transitions and one 2Q transition collectively.

Lastly, we need to comment on the dependence of our results on the computational box size. In general, nuclear spins that are distant from the CS do not affect the spin fluctuations due to dominance of the closed-range nuclear spins, and thereby enable us to work with relatively small-sized computational boxes. Since the proximity of a nuclear spin site to the qubit (i.e., CS) imposes the hf coupling constant, the contribution to the Overhauser field from a nuclear spin site which is separated by a few electron confinement radius is much weaker. Therefore, it is possible to reliably simulate much larger NSBs as long as the CCE-order convergence is assured at the working spinful abundance ratio so that the distanced correlations are quite isolated and are not significantly relayed to the proximate environment in the time frame of the analysis. From a practical point of view, for all of our computations we verified that 1Q and 2Q channel scalograms qualitatively remain intact when we double all of the box dimensions (not shown).

\subsection{Hyperfine axis}

\begin{figure}[tb!]
     \begin{center}
       \includegraphics[width=1\columnwidth]{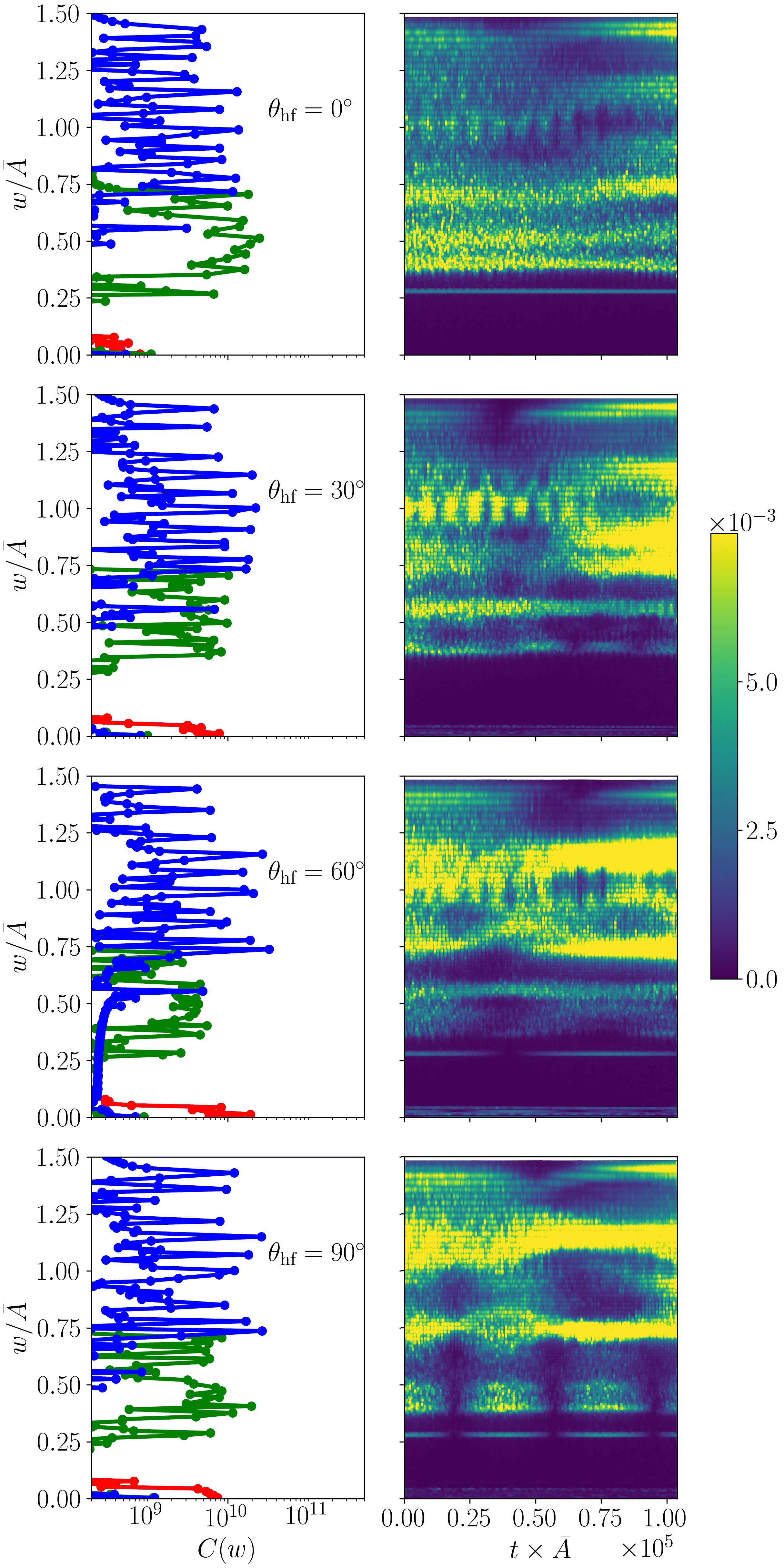}
       \caption{Fourier transforms and SST scalograms of nuclear spin fluctuations under various hf-axis choices for the spin-1/2 baths, $\bar{A} \approx 13.9$~MHz, $\sigma_{\text{hf}} \approx 0.25 \bar{A}$, $N=280$, $\rho= 0.05$, $a_0 = 5.43$~\r{A} (silicon), and $R_c = 2.0 a_0$, calculated via CCE-3. Left column shows the power spectra of two-point correlation functions; right column displays wavelet transforms. Red, green, and blue lines show the coarse-grained contributions of dipolar alphabet $\oB$, $\oC,\oD$ and $\oE$, and $\oF$, respectively.}
       \label{theta_dep}
     \end{center}
 \end{figure}

Now, we quantitatively examine the hf-axis dependence on NSB dynamics. The contributions from distinct $\Delta m_F $ transitions to the power spectra are plotted in the left column of Fig.~\ref{theta_dep} to individually highlight the hf-axis dependencies. This can be simply achieved by manually setting on and off the d-d Hamiltonian terms $\oB$, $\oC,\oD$ and $\oE,\oF$, while retaining $\oA$ since it introduces slight shifts in energy levels without initiating any transitions. The power spectra amplitudes change for each transition channel with respect to hf axis, as it directly determines $\theta_{ij}$ (see Fig.~\ref{crystal}) and, hence, the contribution of each d-d term accordingly. Overall, the spectral overlaps of 0Q, 1Q, and 2Q bands as a function of hf axis are clearly exposed on the left column of Fig.~\ref{theta_dep}. The $\theta_{ij}$ dependence of NSB was also illustrated in Ref.~\cite{witzel06} within the secular approximation by means of Hahn echo decay which confirms our 0Q channel results.

The SSTs are also displayed in the right column of Fig.~\ref{theta_dep}. As corroborated with the power spectra at $\theta_{\text{hf}} = 0 ^\circ$, the main nuclear spin diffusion occurs in the 1Q ($\bar{\omega} = 0.5$) channel. The transitions around  $\bar{\omega} = 0$ are highly suppressed since the $\oB$ term vanishes for the clusters of nearest-neighboring spins. However, as $\theta_{\text{hf}}$ deviates from $0^\circ$ the wavelet transform can capture these nonvanishing transitions. When compared to Fig.~\ref{cce_comp}, due to larger spinful abundance ratio in this case, the 2Q ($\bar{\omega} = 1.0$) channel constitutes complex beating patterns reflecting the physical configuration of second or higher nearest-neighboring spins.

\subsection{Different NSB realizations}

\begin{figure*}
     \begin{center}
       \includegraphics[height=0.48\textheight]{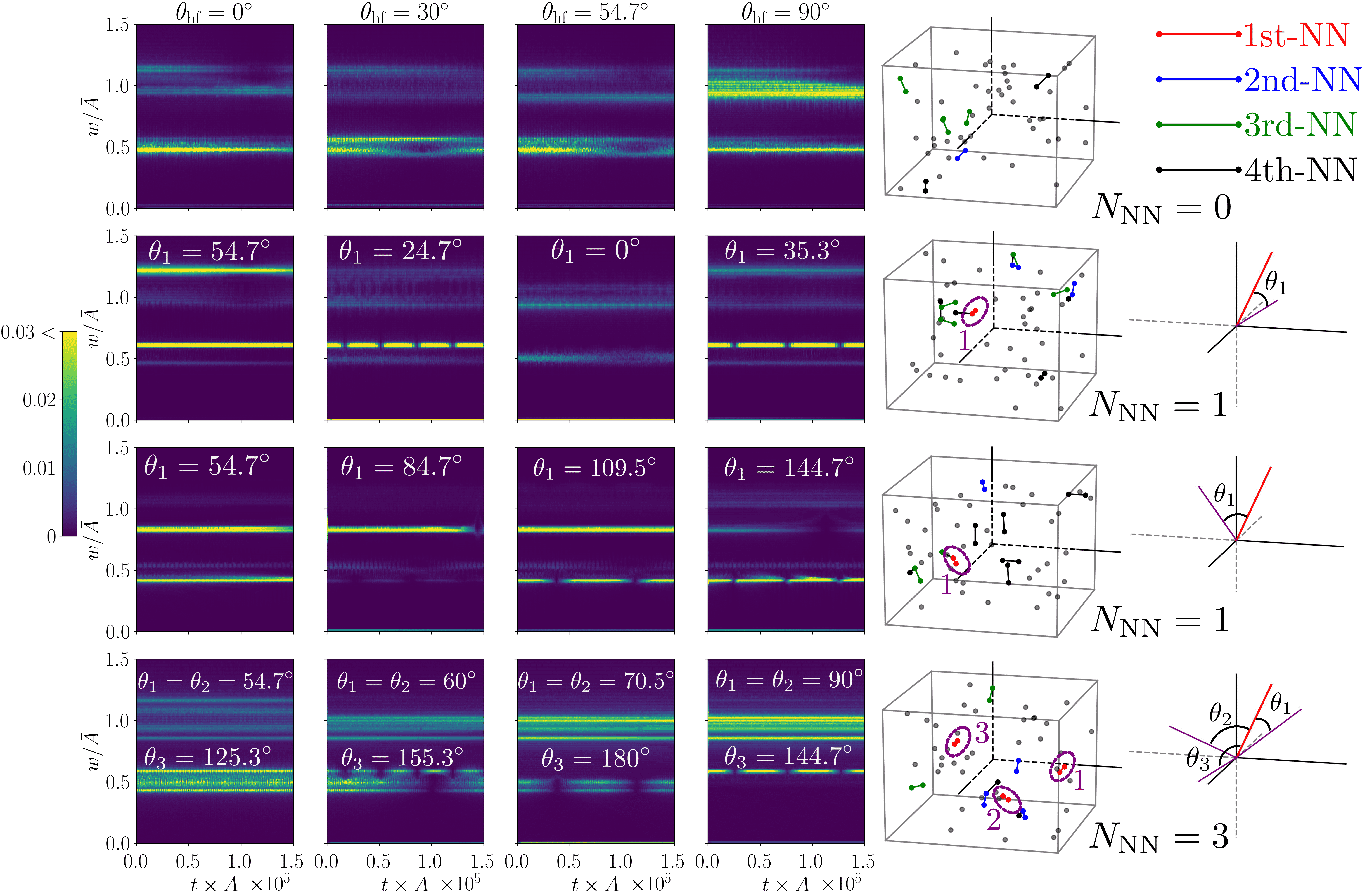}
       \caption{SST of two-point correlation functions of four different bath realizations under various hf axes for spin-1/2 spin environments. Each row represents distinct bath realization with common parameters $N = 56$, $\rho = 0.01$,  $a_0 = 3.567$~\r{A} (diamond), and $R_c = 2.5 a_0$ calculated via CCE-3; the mean value ($\bar{A}$)  and standard deviation ($\sigma_{\text{hf}}$) of hf couplings are around $104$~MHz and $0.12\bar{A}$ ,respectively. Fifth column displays the spatial orientation of nuclear spin environment; the nearest-neighboring spin clusters are encircled and labeled accordingly. While first realization has no nearest-neighboring spin cluster, second and third realization each have one nearest-neighboring spin cluster and fourth realization has three nearest-neighboring spin clusters. Last column shows the alignment of each nearest-neighboring spin cluster and hf axes (red lines) with respect to the crystallographic axes, the angle between displacement vector and hf-axis of interest is labeled as $\theta_i$.}
       \label{realizations}
     \end{center}
 \end{figure*}
We next turn to how a specific spatial distribution of spinful nuclei leaves its imprint on the multiresolution scalogram. As we show below, for dilute NSBs in which the spinful nuclear abundance ratio is small, such as in diamond with a low percentage of the $^{13}$C isotope, this enables us to identify the number of nearest-neighboring spin clusters. Moreover, the angle between the displacement vector that connects two nearest-neighboring spins and the crystallographic axes can also be deduced by simply changing the direction of the hf axis. To demonstrate these, in Fig.~\ref{realizations} each row represents a different NSB realization under various hf axes. In the second column from the right, the spinful nuclei in each realization are designated by dots and neighboring spins are connected with lines and colored with respect to their separation distance. In the very right column of Fig.~\ref{realizations}, the displacement vectors are also drawn to indicate the corresponding angles between hf axis and the displacement vectors.

In the first realization (top row), intentionally there are no nearest-neighboring spin occurrences; the main features are induced by the second-nearest neighbors. The second realization (row) contains one nearest-neighboring spin cluster which is directly captured by the 1Q transition channels of  $\theta_{\text{hf}} = 30^\circ$ and $\theta_{\text{hf}} = 90^\circ$; here there is no beating pattern in the $\theta_{\text{hf}} = 54.7^\circ$ case due to vanishing $\oC$ and $\oD$ terms. Similarly to the second realization, the fourth realization contains a third cluster having identical beat characteristics. This can be seen from the polar angles $\theta_1$ of the second realization and $\theta_3$ of the fourth realization being equal. Moreover, similar beating profiles can be observed from Fig.~\ref{theta_dep} for $\theta_{\text{hf}} = 30^\circ$ and $\theta_{\text{hf}} = 90^\circ$. The nearest-neighboring cluster in the third realization can be distinguished by $\theta_{\text{hf}} = 30^\circ$ which has a slightly higher beating period when compared to the second realization. Finally, in the fourth realization there are three nearest-neighboring spin clusters available. Cluster 1 and cluster 2 have the same orientation with respect to the hf axis, giving rise to a unique fingerprint for $\theta_{\text{hf}} = 30^\circ$. Potentially, this beating profile can be misidentified to be originating from the second-nearest-neighboring spin clusters as in the first realization at the same angle. Fortunately, these two cases can be easily discriminated by reorienting the hf axis to [111] the direction.

Owing to the much sharper spectra with SSTs, it is possible to gain information about the distance between nearest-neighboring clusters and CS. Since the $A_i$'s are distributed according to spatial distance to the CS, as nearest-neighboring spin clusters get closer to the qubit the mean values of nearest-neighboring clusters grow, and the beating patterns shift to higher frequencies. This can be seen in the fourth realization of Fig.~\ref{realizations}, in comparison to the beating positions in the second and third realization along the frequency axis.

 \subsection{High magnetic field regime} \label{highBfieldregime}

 \begin{figure}[b!]
      \begin{center}
       \includegraphics[width=1\columnwidth]{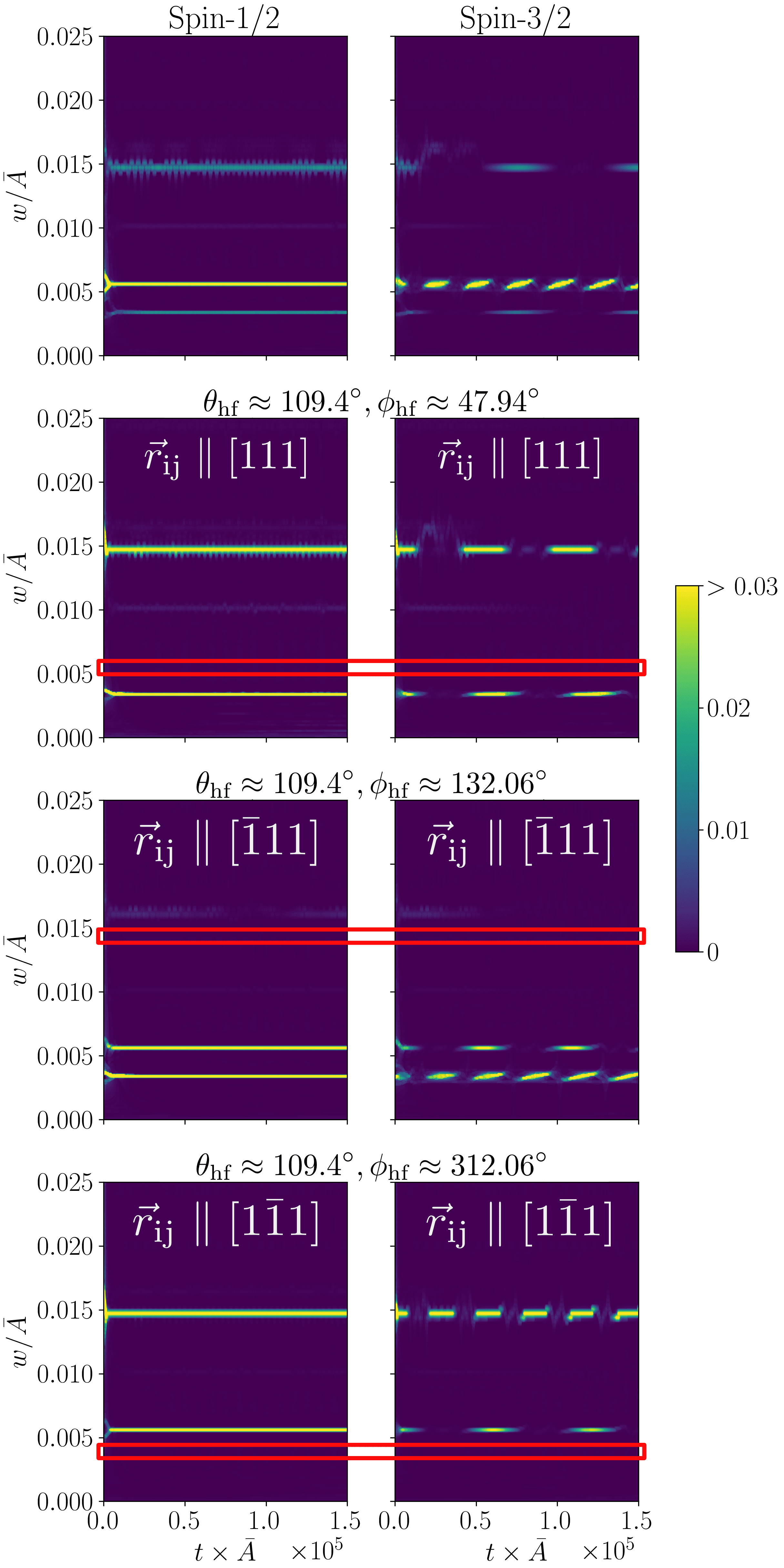}
        \caption{ SST scalograms of spin fluctuations of fourth realization of Fig.~\ref{realizations} in high magnetic field regime for spin-1/2 (left) and spin-3/2 (right) baths,  $\bar{A} \approx 104.7$~MHz, $\sigma_{\text{hf}} \approx 0.12 \bar{A}$, $N=56$, $\rho= 0.01$, $a_0 = 3.567$~\r{A} (diamond), and $R_c = 2.5 a_0$, calculated via CCE-3. Upper row shows time-frequency analysis for hf axis in [111] direction, while other rows display the SST for various hf axes in which contribution from each bond axis is eliminated successively to identify the orientation of nearest-neighboring spin clusters as marked in red rectangles.}
        \label{high_B}
      \end{center}
  \end{figure}

We proceed with the fourth realization in Fig.~\ref{realizations} in which there are three nearest-neighboring spin clusters, and compare in Fig.~\ref{high_B} the SST for spin-1/2 (left) and spin-3/2 (right) NSBs under specific hf axes.
For the spin-1/2 bath, only a single 0Q  transition is available which occurs between two $|m_F=0\rangle$ states. Hence, there is no beating pattern even if $|m_F=0\rangle$ is perturbed through the $\oA$ term. On the other hand, for the spin-3/2 NSB multiple 0Q transitions are available. Thus, the collective detunings of 0Q transitions yield the resultant beating pattern as in the first row of Fig.~\ref{high_B}.

Even more information on the spatial distribution of the spinful nuclei can be extracted. For instance, the number of nearest-neighboring spins and the alignment of displacement vectors can also be deduced. This is achieved by the following procedure: \textit{i}) Obtain the SST for both [001] and [111] hf-axis directions, noting that when the hf axis is aligned in the [001] direction, contributions from first-nearest-neighboring spin clusters are negated, whereas in the [111] direction all nearest-neighboring spins contribute to the spin noise, so that all nearest-neighboring spins are located as horizontal stripes (spin-1/2) or beating (spin-3/2); see the first row of Fig.~\ref{high_B} around $w/\bar{A}=0.004$, $w/\bar{A}=0.006$, $w/\bar{A}=0.015$. \textit{ii}) Arrange the hf axis so that the angle between one of the possible bond axes and the direction of the hf axis becomes $54.7^\circ$. This removes a horizontal stripe(s) or beating pattern(s) in the SST if there is any nearest-neighboring cluster along the relevant bond axis, as marked by red rectangular frames in Fig.~\ref{high_B}. \textit{iii}) Repeating the step (\textit{ii}) for each bond axis until capturing the alignment of  all nearest-neighboring spin clusters as indicated in last  three rows of Fig.~\ref{high_B}. It is important to note that unlike the low-field regime, the distance between CS and nearest-neighboring spins is not directly implied by the frequency axis position of stripes or beating patterns in time-frequency analysis since it depends on the detuning of hf couplings of nearest-neighboring spins.

\subsection{Noise resilience of coherence beats in the wavelet scalograms}
Our model Hamiltonian (Eq.~(\ref{Genel Ham})) can be enriched with more interactions present in realistic cases, such as the electric quadrupolar coupling for $I\ge 1$ nuclei \cite{bulutay12} which allows for co-flips with the CS \cite{hogele12}, or hf-mediated long-range interactions \cite{cywinskiprb09}. To have a glimpse on the performance of SST under more adverse conditions than Eq.~(\ref{Genel Ham}), we would like to include a background noise, albeit in a rather primitive way.
A well-established model for this purpose is the random telegraph noise (RTN) that accounts for classical sources as in the charge noise \cite{szankowski17}. We introduce RTN to the Hamiltonian in Eq.~(\ref{effective_Hamiltonian}) as a stochastic time-dependent term
 \begin{equation}
 	\Xi(t) = \sum_{i=1}^N \nu_i\xi_i(t)I^z_i,
 \end{equation}
where $i$ is the nuclear spin index, with the amplitude $\nu_i$ being chosen from a Gaussian probability distribution with mean value $\bar{\nu}$ and standard deviation $\bar{\nu}/4$. $\xi_i$ is the so-called switching function with the rate $\Gamma$ taking binary values $\pm 1$ so that each nuclear spin switches once at a random instance within a time interval $\Delta t=1/\Gamma$. Following the previous studies of this model, we also define the dimensionless parameter $\eta = \bar{\nu}/\Gamma$ which quantifies weak ($\eta\ll 1$) and strong ($\eta\gg 1$) coupling regimes \cite{ramon15}.

\begin{figure}[tb!]
		 \begin{center}
			\includegraphics[width=1\columnwidth]{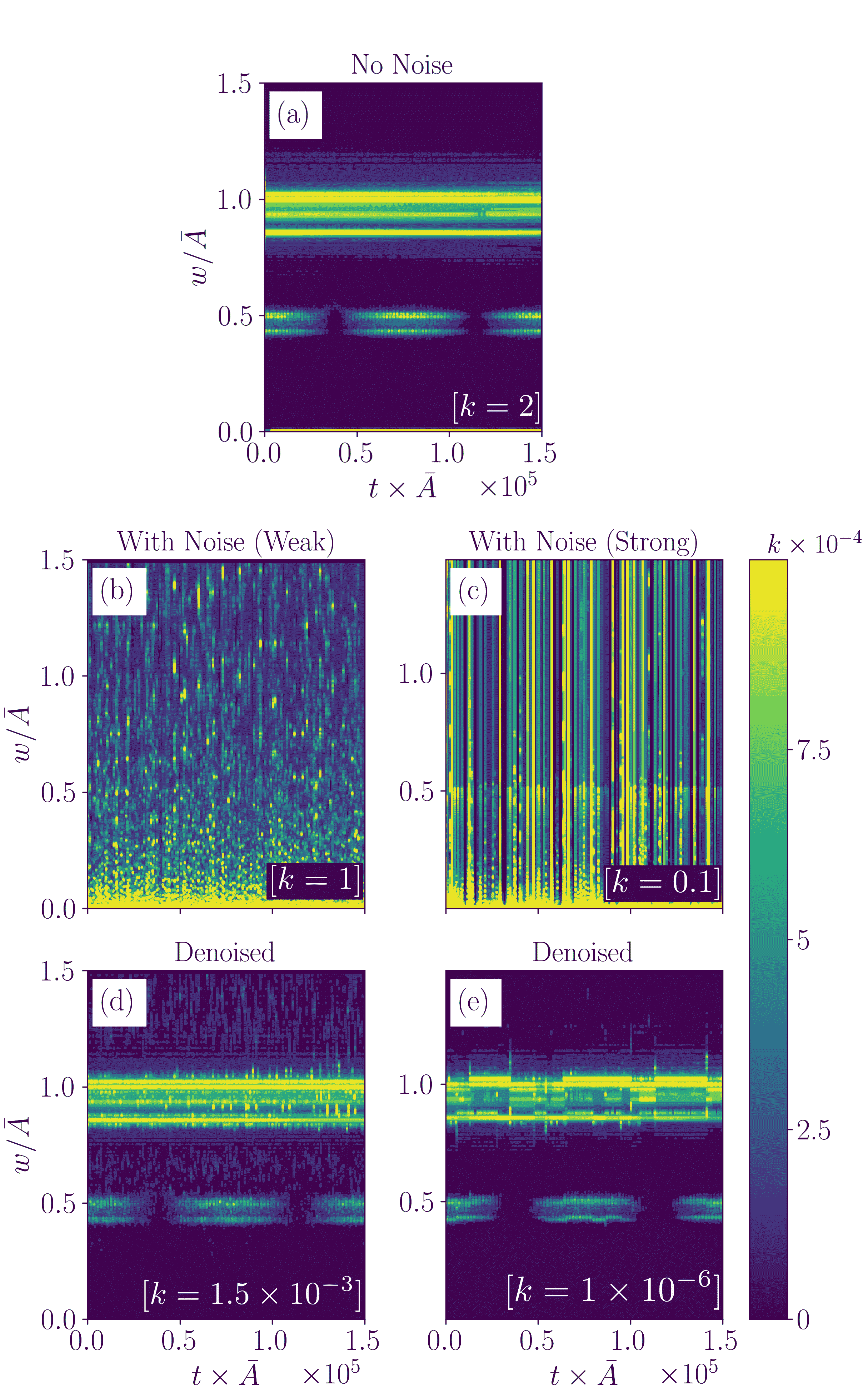}
			 \caption{ Comparison of the SST scalogram of the fourth realization of Fig.~\ref{realizations}, here panel (a), with the weakly ($\eta \approx 0.01$) in (b), and strongly ($\eta \approx 100$) in (c), RTN regimes. The mean fluctuator coupling strength, $\bar{\nu}$, and switching frequency, $\Gamma$, are chosen to be $1 \times 10^{-2} \bar{A}$ ($1 \times 10^{-5} \bar{A}$) and $10$~kHz ($100$~kHz) for strong (weak) coupling regimes, respectively. Panels (d) and (e) display the denoised versions of (b) and (c), respectively via SWT. $k$ is the scaling factor of the color bar among all panels. $\theta_{\text{hf}}  \approx 54.7^\circ$; other details and parameters are the same as Fig.~\ref{realizations}.}
			 \label{rtn}
		 \end{center}
 \end{figure}

Figure~\ref{rtn} contrasts the noiseless SST of $\bar{C}(t)$ reproduced from the fourth realization of Fig.~\ref{realizations} with those under noise within the weak ($\eta \approx 0.01$), and strong ($\eta \approx 100$) coupling regimes. As a side note, the weakly coupled case for RTN is known to satisfy the Gaussian limit \cite{cywinski08, ramon15, szankowski17}. From Fig.~\ref{rtn}~(b) we observe that the beating in 1Q regions become completely invisible, even under the weakly coupled noise. Fortunately, it is possible to denoise the nuclear spin correlations from RTN, and recover the beating profile significantly, as shown in  Fig.~\ref{rtn}~(d). This is achieved by employing a discrete stationary wavelet transform (SWT) by means of careful thresholding of detailed coefficients at required decomposition levels \cite{nason95}. Specific to RTN, the Haar wavelet is a suitable option to recover nuclear spin correlations. We should note that while the beating patterns are successfully recovered in the weak coupling regime for which $\Gamma$ is chosen to be in the hundred kilohertz range, this becomes much harder using conventional wavelet-based denoising techniques in the megahertz range for our exemplary parameter sets.

The strong coupling regime ($\eta\approx100$) behavior is displayed in Fig.~\ref{rtn}~(c), where the stochastic two-level fluctuations are distinguishable as vertical stripes due to low RTN rate. The remnants of strongly coupled noise somewhat persist in the denoised scalogram as indicated in Fig.~\ref{rtn}~(e). Overall, though not as clean as the weak coupling case, filtering out the strongly coupled background noise results in an acceptable scalogram. Nonetheless, given the simplistic nature of our Hamiltonian and noise models, these can only be taken as the first steps toward application of wavelet analysis in semiconductor spin coherence spectroscopy.

\section{Discussions}
%\subsection{Possible applications of wavelet analysis}

In this section we would like to put into perspective our study in the light of two recent reconstruction accomplishments enabled by nuclear spin ensemble coherence \cite{abobeih19,gangloff21}. Strikingly, the spin baths as well as the goals in these experiments are remarkably different from each other. In one case, Gangloff and co-workers reconstructed species-resolved spin states of an InGaAs self-assembled quantum dot involving tens of thousands of spinful quadrupolar nuclei \cite{gangloff21}. This makes up a dense NSB which is polarized via dragging \cite{hogele12}, and through magnon spectroscopy they verified a non-thermal, correlated nuclear state. At this point we should note that our work is based on an unpolarized and dilute spin bath. Actually these choices arise from the CCE technique \cite{yang08, yang09, ma15}. Namely, the unpolarized (i.e., at infinite temperature) NSB stems from avoiding the cross-correlation elements of the two-point correlation function \cite{ma15,yang17}, and spatially dilute NSB hinges on the convergence issues for both large and dense spin baths. Notwithstanding, the wavelet analysis presented here is by no means limited to these cases. As a matter of fact, from an experimental standpoint thermal ensembles are usually undesirable due to reduced electron, i.e., CS, coherence times \cite{urbaszek13}. Thus, the use of wavelets in polarized and/or dense NSB to reveal quantum coherence signatures in experimental data can be indeed practical and interesting \cite{gangloff21}.

In another work, Abobeih \textit{et al.} succeeded in atomic-scale imaging using a single nitrogen vacancy center as a quantum sensor, and spatially reconstructed a model system of 27 coupled $^{13}$C nuclear spins in diamond \cite{abobeih19}. Their approach relies on the successful isolation of a probe spin from environmental nuclear spin noise and recoupling to a target spin, by means of nuclear-nuclear double-resonance spectroscopy. This is certainly a sophisticated procedure, and it necessitates several radio frequency pulse sequence sets. We believe that as a complementary tool, the wavelet-based time-frequency analysis can be instrumental in nanoscale NMR imaging of proximal cluster spins by capturing their unique beating patterns in the SST scalogram, as outlined in this work.

\section{Conclusions}
The coherent dipolar dynamics of a NSB hf-coupled to a CS inherits crucial information about the spatial
distribution of the spinful nuclei. In this work we demonstrate how to extract these from the two-point correlation
function by postprocessing with wavelet analysis, in particular using SST for enhanced spectral resolution.
Foremost, probing the existence of beating patterns, which is in principle possible even
for a thermal NSB, is indicative of quantum coherence within the NSB.
The quadrupolar nuclei, if present within the bath, exhibit themselves as beating residing in the 0Q channel.
By harnessing the choice of hf axis in the low-magnetic-field regime, it is possible to identify the spatial distance to CS and displacement vector alignment of the spinful nearest-neighboring nuclear sites.
Additionally, in the presence of large magnetic fields in which the secular terms of the d-d alphabet are considered only, multiresolution analysis is again able to capture nearest-neighboring features. Lastly, in the more realistic case of coherence beating being buried by environmental noise,
wavelet domain thresholding techniques are very effective in denoising the scalograms. We hope that these promising findings as a first step can motivate further theoretical and experimental investigations of the wavelet analysis in diagnosing the nuclear spin environment.

\begin{acknowledgments}
This work was funded by T\"urkiye Bilimsel ve Teknolojik Ara\c{s}tirma Kurumu
(TUBITAK) under Project No.~116F075.
The numerical calculations reported in this paper were partially performed at Türkiye Bilimsel ve Teknolojik Araştirma Kurumu (TUBITAK) ULAKBIM, High Performance, and Grid Computing Center (TRUBA resources).
\end{acknowledgments}


\begin{thebibliography}{92}%
\makeatletter
\providecommand \@ifxundefined [1]{%
 \@ifx{#1\undefined}
}%
\providecommand \@ifnum [1]{%
 \ifnum #1\expandafter \@firstoftwo
 \else \expandafter \@secondoftwo
 \fi
}%
\providecommand \@ifx [1]{%
 \ifx #1\expandafter \@firstoftwo
 \else \expandafter \@secondoftwo
 \fi
}%
\providecommand \natexlab [1]{#1}%
\providecommand \enquote  [1]{``#1''}%
\providecommand \bibnamefont  [1]{#1}%
\providecommand \bibfnamefont [1]{#1}%
\providecommand \citenamefont [1]{#1}%
\providecommand \href@noop [0]{\@secondoftwo}%
\providecommand \href [0]{\begingroup \@sanitize@url \@href}%
\providecommand \@href[1]{\@@startlink{#1}\@@href}%
\providecommand \@@href[1]{\endgroup#1\@@endlink}%
\providecommand \@sanitize@url [0]{\catcode `\\12\catcode `\$12\catcode
  `\&12\catcode `\#12\catcode `\^12\catcode `\_12\catcode `\%12\relax}%
\providecommand \@@startlink[1]{}%
\providecommand \@@endlink[0]{}%
\providecommand \url  [0]{\begingroup\@sanitize@url \@url }%
\providecommand \@url [1]{\endgroup\@href {#1}{\urlprefix }}%
\providecommand \urlprefix  [0]{URL }%
\providecommand \Eprint [0]{\href }%
\providecommand \doibase [0]{https://doi.org/}%
\providecommand \selectlanguage [0]{\@gobble}%
\providecommand \bibinfo  [0]{\@secondoftwo}%
\providecommand \bibfield  [0]{\@secondoftwo}%
\providecommand \translation [1]{[#1]}%
\providecommand \BibitemOpen [0]{}%
\providecommand \bibitemStop [0]{}%
\providecommand \bibitemNoStop [0]{.\EOS\space}%
\providecommand \EOS [0]{\spacefactor3000\relax}%
\providecommand \BibitemShut  [1]{\csname bibitem#1\endcsname}%
\let\auto@bib@innerbib\@empty
%</preamble>
\bibitem [{\citenamefont {Awschalom}\ \emph {et~al.}(2013)\citenamefont
  {Awschalom}, \citenamefont {Bassett}, \citenamefont {Dzurak}, \citenamefont
  {Hu},\ and\ \citenamefont {Petta}}]{awschalom13}%
  \BibitemOpen
  \bibfield  {author} {\bibinfo {author} {\bibfnamefont {D.~D.}\ \bibnamefont
  {Awschalom}}, \bibinfo {author} {\bibfnamefont {L.~C.}\ \bibnamefont
  {Bassett}}, \bibinfo {author} {\bibfnamefont {A.~S.}\ \bibnamefont {Dzurak}},
  \bibinfo {author} {\bibfnamefont {E.~L.}\ \bibnamefont {Hu}},\ and\ \bibinfo
  {author} {\bibfnamefont {J.~R.}\ \bibnamefont {Petta}},\ }\bibfield  {title}
  {\bibinfo {title} {Quantum spintronics: Engineering and manipulating
  atom-like spins in semiconductors},\ }\href
  {https://doi.org/10.1126/science.1231364} {\bibfield  {journal} {\bibinfo
  {journal} {Science}\ }\textbf {\bibinfo {volume} {339}},\ \bibinfo {pages}
  {1174} (\bibinfo {year} {2013})},\ \Eprint
  {https://arxiv.org/abs/https://science.sciencemag.org/content/339/6124/1174.full.pdf}
  {https://science.sciencemag.org/content/339/6124/1174.full.pdf} \BibitemShut
  {NoStop}%
\bibitem [{\citenamefont {Chatterjee}\ \emph {et~al.}(2021)\citenamefont
  {Chatterjee}, \citenamefont {Stevenson}, \citenamefont {De~Franceschi},
  \citenamefont {Morello}, \citenamefont {de~Leon},\ and\ \citenamefont
  {Kuemmeth}}]{chatterjee21}%
  \BibitemOpen
  \bibfield  {author} {\bibinfo {author} {\bibfnamefont {A.}~\bibnamefont
  {Chatterjee}}, \bibinfo {author} {\bibfnamefont {P.}~\bibnamefont
  {Stevenson}}, \bibinfo {author} {\bibfnamefont {S.}~\bibnamefont
  {De~Franceschi}}, \bibinfo {author} {\bibfnamefont {A.}~\bibnamefont
  {Morello}}, \bibinfo {author} {\bibfnamefont {N.~P.}\ \bibnamefont
  {de~Leon}},\ and\ \bibinfo {author} {\bibfnamefont {F.}~\bibnamefont
  {Kuemmeth}},\ }\bibfield  {title} {\bibinfo {title} {Semiconductor qubits in
  practice},\ }\href {https://doi.org/10.1038/s42254-021-00283-9} {\bibfield
  {journal} {\bibinfo  {journal} {Nature Reviews Physics}\ }\textbf {\bibinfo
  {volume} {3}},\ \bibinfo {pages} {157} (\bibinfo {year} {2021})}\BibitemShut
  {NoStop}%
\bibitem [{\citenamefont {Wolfowicz}\ \emph {et~al.}(2013)\citenamefont
  {Wolfowicz}, \citenamefont {Tyryshkin}, \citenamefont {George}, \citenamefont
  {Riemann}, \citenamefont {Abrosimov}, \citenamefont {Becker}, \citenamefont
  {Pohl}, \citenamefont {Thewalt}, \citenamefont {Lyon},\ and\ \citenamefont
  {Morton}}]{wolfowicz13}%
  \BibitemOpen
  \bibfield  {author} {\bibinfo {author} {\bibfnamefont {G.}~\bibnamefont
  {Wolfowicz}}, \bibinfo {author} {\bibfnamefont {A.~M.}\ \bibnamefont
  {Tyryshkin}}, \bibinfo {author} {\bibfnamefont {R.~E.}\ \bibnamefont
  {George}}, \bibinfo {author} {\bibfnamefont {H.}~\bibnamefont {Riemann}},
  \bibinfo {author} {\bibfnamefont {N.~V.}\ \bibnamefont {Abrosimov}}, \bibinfo
  {author} {\bibfnamefont {P.}~\bibnamefont {Becker}}, \bibinfo {author}
  {\bibfnamefont {H.-J.}\ \bibnamefont {Pohl}}, \bibinfo {author}
  {\bibfnamefont {M.~L.~W.}\ \bibnamefont {Thewalt}}, \bibinfo {author}
  {\bibfnamefont {S.~A.}\ \bibnamefont {Lyon}},\ and\ \bibinfo {author}
  {\bibfnamefont {J.~J.~L.}\ \bibnamefont {Morton}},\ }\bibfield  {title}
  {\bibinfo {title} {Atomic clock transitions in silicon-based spin qubits},\
  }\href {https://doi.org/10.1038/nnano.2013.117} {\bibfield  {journal}
  {\bibinfo  {journal} {Nature Nanotechnology}\ }\textbf {\bibinfo {volume}
  {8}},\ \bibinfo {pages} {561} (\bibinfo {year} {2013})}\BibitemShut {NoStop}%
\bibitem [{\citenamefont {Cai}\ \emph {et~al.}(2013)\citenamefont {Cai},
  \citenamefont {Retzker}, \citenamefont {Jelezko},\ and\ \citenamefont
  {Plenio}}]{cai13}%
  \BibitemOpen
  \bibfield  {author} {\bibinfo {author} {\bibfnamefont {J.}~\bibnamefont
  {Cai}}, \bibinfo {author} {\bibfnamefont {A.}~\bibnamefont {Retzker}},
  \bibinfo {author} {\bibfnamefont {F.}~\bibnamefont {Jelezko}},\ and\ \bibinfo
  {author} {\bibfnamefont {M.~B.}\ \bibnamefont {Plenio}},\ }\bibfield  {title}
  {\bibinfo {title} {A large-scale quantum simulator on a diamond surface at
  room temperature},\ }\href {https://doi.org/10.1038/nphys2519} {\bibfield
  {journal} {\bibinfo  {journal} {Nature Physics}\ }\textbf {\bibinfo {volume}
  {9}},\ \bibinfo {pages} {168} (\bibinfo {year} {2013})}\BibitemShut {NoStop}%
\bibitem [{\citenamefont {Choi}\ \emph {et~al.}(2017)\citenamefont {Choi},
  \citenamefont {Choi}, \citenamefont {Landig}, \citenamefont {Kucsko},
  \citenamefont {Zhou}, \citenamefont {Isoya}, \citenamefont {Jelezko},
  \citenamefont {Onoda}, \citenamefont {Sumiya}, \citenamefont {Khemani},
  \citenamefont {von Keyserlingk}, \citenamefont {Yao}, \citenamefont
  {Demler},\ and\ \citenamefont {Lukin}}]{choi17}%
  \BibitemOpen
  \bibfield  {author} {\bibinfo {author} {\bibfnamefont {S.}~\bibnamefont
  {Choi}}, \bibinfo {author} {\bibfnamefont {J.}~\bibnamefont {Choi}}, \bibinfo
  {author} {\bibfnamefont {R.}~\bibnamefont {Landig}}, \bibinfo {author}
  {\bibfnamefont {G.}~\bibnamefont {Kucsko}}, \bibinfo {author} {\bibfnamefont
  {H.}~\bibnamefont {Zhou}}, \bibinfo {author} {\bibfnamefont {J.}~\bibnamefont
  {Isoya}}, \bibinfo {author} {\bibfnamefont {F.}~\bibnamefont {Jelezko}},
  \bibinfo {author} {\bibfnamefont {S.}~\bibnamefont {Onoda}}, \bibinfo
  {author} {\bibfnamefont {H.}~\bibnamefont {Sumiya}}, \bibinfo {author}
  {\bibfnamefont {V.}~\bibnamefont {Khemani}}, \bibinfo {author} {\bibfnamefont
  {C.}~\bibnamefont {von Keyserlingk}}, \bibinfo {author} {\bibfnamefont
  {N.~Y.}\ \bibnamefont {Yao}}, \bibinfo {author} {\bibfnamefont
  {E.}~\bibnamefont {Demler}},\ and\ \bibinfo {author} {\bibfnamefont {M.~D.}\
  \bibnamefont {Lukin}},\ }\bibfield  {title} {\bibinfo {title} {Observation of
  discrete time-crystalline order in a disordered dipolar many-body system},\
  }\href {https://doi.org/10.1038/nature21426} {\bibfield  {journal} {\bibinfo
  {journal} {Nature}\ }\textbf {\bibinfo {volume} {543}},\ \bibinfo {pages}
  {221} (\bibinfo {year} {2017})}\BibitemShut {NoStop}%
\bibitem [{\citenamefont {Wang}\ \emph {et~al.}(2019)\citenamefont {Wang},
  \citenamefont {Chen}, \citenamefont {Peng}, \citenamefont {Wrachtrup},\ and\
  \citenamefont {Liu}}]{wang19}%
  \BibitemOpen
  \bibfield  {author} {\bibinfo {author} {\bibfnamefont {P.}~\bibnamefont
  {Wang}}, \bibinfo {author} {\bibfnamefont {C.}~\bibnamefont {Chen}}, \bibinfo
  {author} {\bibfnamefont {X.}~\bibnamefont {Peng}}, \bibinfo {author}
  {\bibfnamefont {J.}~\bibnamefont {Wrachtrup}},\ and\ \bibinfo {author}
  {\bibfnamefont {R.-B.}\ \bibnamefont {Liu}},\ }\bibfield  {title} {\bibinfo
  {title} {Characterization of arbitrary-order correlations in quantum baths by
  weak measurement},\ }\href {https://doi.org/10.1103/PhysRevLett.123.050603}
  {\bibfield  {journal} {\bibinfo  {journal} {Phys. Rev. Lett.}\ }\textbf
  {\bibinfo {volume} {123}},\ \bibinfo {pages} {050603} (\bibinfo {year}
  {2019})}\BibitemShut {NoStop}%
\bibitem [{\citenamefont {Jackson}\ \emph {et~al.}(2021)\citenamefont
  {Jackson}, \citenamefont {Gangloff}, \citenamefont {Bodey}, \citenamefont
  {Zaporski}, \citenamefont {Bachorz}, \citenamefont {Clarke}, \citenamefont
  {Hugues}, \citenamefont {Le~Gall},\ and\ \citenamefont
  {Atat{\"u}re}}]{jackson21}%
  \BibitemOpen
  \bibfield  {author} {\bibinfo {author} {\bibfnamefont {D.~M.}\ \bibnamefont
  {Jackson}}, \bibinfo {author} {\bibfnamefont {D.~A.}\ \bibnamefont
  {Gangloff}}, \bibinfo {author} {\bibfnamefont {J.~H.}\ \bibnamefont {Bodey}},
  \bibinfo {author} {\bibfnamefont {L.}~\bibnamefont {Zaporski}}, \bibinfo
  {author} {\bibfnamefont {C.}~\bibnamefont {Bachorz}}, \bibinfo {author}
  {\bibfnamefont {E.}~\bibnamefont {Clarke}}, \bibinfo {author} {\bibfnamefont
  {M.}~\bibnamefont {Hugues}}, \bibinfo {author} {\bibfnamefont
  {C.}~\bibnamefont {Le~Gall}},\ and\ \bibinfo {author} {\bibfnamefont
  {M.}~\bibnamefont {Atat{\"u}re}},\ }\bibfield  {title} {\bibinfo {title}
  {Quantum sensing of a coherent single spin excitation in a nuclear
  ensemble},\ }\href {https://doi.org/10.1038/s41567-020-01161-4} {\bibfield
  {journal} {\bibinfo  {journal} {Nature Physics}\ }\textbf {\bibinfo {volume}
  {17}},\ \bibinfo {pages} {585} (\bibinfo {year} {2021})}\BibitemShut
  {NoStop}%
\bibitem [{\citenamefont {Khaetskii}\ \emph {et~al.}(2002)\citenamefont
  {Khaetskii}, \citenamefont {Loss},\ and\ \citenamefont
  {Glazman}}]{khaetskii02}%
  \BibitemOpen
  \bibfield  {author} {\bibinfo {author} {\bibfnamefont {A.~V.}\ \bibnamefont
  {Khaetskii}}, \bibinfo {author} {\bibfnamefont {D.}~\bibnamefont {Loss}},\
  and\ \bibinfo {author} {\bibfnamefont {L.}~\bibnamefont {Glazman}},\
  }\bibfield  {title} {\bibinfo {title} {Electron spin decoherence in quantum
  dots due to interaction with nuclei},\ }\href
  {https://doi.org/10.1103/PhysRevLett.88.186802} {\bibfield  {journal}
  {\bibinfo  {journal} {Phys. Rev. Lett.}\ }\textbf {\bibinfo {volume} {88}},\
  \bibinfo {pages} {186802} (\bibinfo {year} {2002})}\BibitemShut {NoStop}%
\bibitem [{\citenamefont {Merkulov}\ \emph {et~al.}(2002)\citenamefont
  {Merkulov}, \citenamefont {Efros},\ and\ \citenamefont {Rosen}}]{merkulov02}%
  \BibitemOpen
  \bibfield  {author} {\bibinfo {author} {\bibfnamefont {I.~A.}\ \bibnamefont
  {Merkulov}}, \bibinfo {author} {\bibfnamefont {A.~L.}\ \bibnamefont
  {Efros}},\ and\ \bibinfo {author} {\bibfnamefont {M.}~\bibnamefont {Rosen}},\
  }\bibfield  {title} {\bibinfo {title} {Electron spin relaxation by nuclei in
  semiconductor quantum dots},\ }\href
  {https://doi.org/10.1103/PhysRevB.65.205309} {\bibfield  {journal} {\bibinfo
  {journal} {Phys. Rev. B}\ }\textbf {\bibinfo {volume} {65}},\ \bibinfo
  {pages} {205309} (\bibinfo {year} {2002})}\BibitemShut {NoStop}%
\bibitem [{\citenamefont {Coish}\ and\ \citenamefont {Loss}(2004)}]{coish04}%
  \BibitemOpen
  \bibfield  {author} {\bibinfo {author} {\bibfnamefont {W.~A.}\ \bibnamefont
  {Coish}}\ and\ \bibinfo {author} {\bibfnamefont {D.}~\bibnamefont {Loss}},\
  }\bibfield  {title} {\bibinfo {title} {Hyperfine interaction in a quantum
  dot: Non-markovian electron spin dynamics},\ }\href
  {https://doi.org/10.1103/PhysRevB.70.195340} {\bibfield  {journal} {\bibinfo
  {journal} {Phys. Rev. B}\ }\textbf {\bibinfo {volume} {70}},\ \bibinfo
  {pages} {195340} (\bibinfo {year} {2004})}\BibitemShut {NoStop}%
\bibitem [{\citenamefont {Cywi\ifmmode~\acute{n}\else \'{n}\fi{}ski}\ \emph
  {et~al.}(2009{\natexlab{a}})\citenamefont {Cywi\ifmmode~\acute{n}\else
  \'{n}\fi{}ski}, \citenamefont {Witzel},\ and\ \citenamefont
  {Das~Sarma}}]{cywinskiprl09}%
  \BibitemOpen
  \bibfield  {author} {\bibinfo {author} {\bibfnamefont {L.}~\bibnamefont
  {Cywi\ifmmode~\acute{n}\else \'{n}\fi{}ski}}, \bibinfo {author}
  {\bibfnamefont {W.~M.}\ \bibnamefont {Witzel}},\ and\ \bibinfo {author}
  {\bibfnamefont {S.}~\bibnamefont {Das~Sarma}},\ }\bibfield  {title} {\bibinfo
  {title} {Electron spin dephasing due to hyperfine interactions with a nuclear
  spin bath},\ }\href {https://doi.org/10.1103/PhysRevLett.102.057601}
  {\bibfield  {journal} {\bibinfo  {journal} {Phys. Rev. Lett.}\ }\textbf
  {\bibinfo {volume} {102}},\ \bibinfo {pages} {057601} (\bibinfo {year}
  {2009}{\natexlab{a}})}\BibitemShut {NoStop}%
\bibitem [{\citenamefont {Ma}\ \emph {et~al.}(2014)\citenamefont {Ma},
  \citenamefont {Wolfowicz}, \citenamefont {Zhao}, \citenamefont {Li},
  \citenamefont {Morton},\ and\ \citenamefont {Liu}}]{ma14}%
  \BibitemOpen
  \bibfield  {author} {\bibinfo {author} {\bibfnamefont {W.-L.}\ \bibnamefont
  {Ma}}, \bibinfo {author} {\bibfnamefont {G.}~\bibnamefont {Wolfowicz}},
  \bibinfo {author} {\bibfnamefont {N.}~\bibnamefont {Zhao}}, \bibinfo {author}
  {\bibfnamefont {S.-S.}\ \bibnamefont {Li}}, \bibinfo {author} {\bibfnamefont
  {J.~J.~L.}\ \bibnamefont {Morton}},\ and\ \bibinfo {author} {\bibfnamefont
  {R.-B.}\ \bibnamefont {Liu}},\ }\bibfield  {title} {\bibinfo {title}
  {Uncovering many-body correlations in nanoscale nuclear spin baths by central
  spin decoherence},\ }\href {https://doi.org/10.1038/ncomms5822} {\bibfield
  {journal} {\bibinfo  {journal} {Nature Communications}\ }\textbf {\bibinfo
  {volume} {5}},\ \bibinfo {pages} {4822} (\bibinfo {year} {2014})}\BibitemShut
  {NoStop}%
\bibitem [{\citenamefont {Zhang}\ \emph {et~al.}(2020)\citenamefont {Zhang},
  \citenamefont {Ma},\ and\ \citenamefont {Liu}}]{zhang20}%
  \BibitemOpen
  \bibfield  {author} {\bibinfo {author} {\bibfnamefont {G.-L.}\ \bibnamefont
  {Zhang}}, \bibinfo {author} {\bibfnamefont {W.-L.}\ \bibnamefont {Ma}},\ and\
  \bibinfo {author} {\bibfnamefont {R.-B.}\ \bibnamefont {Liu}},\ }\bibfield
  {title} {\bibinfo {title} {Cluster correlation expansion for studying
  decoherence of clock transitions in spin baths},\ }\href
  {https://doi.org/10.1103/PhysRevB.102.245303} {\bibfield  {journal} {\bibinfo
   {journal} {Phys. Rev. B}\ }\textbf {\bibinfo {volume} {102}},\ \bibinfo
  {pages} {245303} (\bibinfo {year} {2020})}\BibitemShut {NoStop}%
\bibitem [{\citenamefont {Hanson}\ \emph {et~al.}(2008)\citenamefont {Hanson},
  \citenamefont {Dobrovitski}, \citenamefont {Feiguin}, \citenamefont {Gywat},\
  and\ \citenamefont {Awschalom}}]{hanson08}%
  \BibitemOpen
  \bibfield  {author} {\bibinfo {author} {\bibfnamefont {R.}~\bibnamefont
  {Hanson}}, \bibinfo {author} {\bibfnamefont {V.~V.}\ \bibnamefont
  {Dobrovitski}}, \bibinfo {author} {\bibfnamefont {A.~E.}\ \bibnamefont
  {Feiguin}}, \bibinfo {author} {\bibfnamefont {O.}~\bibnamefont {Gywat}},\
  and\ \bibinfo {author} {\bibfnamefont {D.~D.}\ \bibnamefont {Awschalom}},\
  }\bibfield  {title} {\bibinfo {title} {Coherent dynamics of a single spin
  interacting with an adjustable spin bath},\ }\href
  {https://doi.org/10.1126/science.1155400} {\bibfield  {journal} {\bibinfo
  {journal} {Science}\ }\textbf {\bibinfo {volume} {320}},\ \bibinfo {pages}
  {352} (\bibinfo {year} {2008})}\BibitemShut {NoStop}%
\bibitem [{\citenamefont {{Gaudin, M.}}(1976)}]{gaudin76}%
  \BibitemOpen
  \bibfield  {author} {\bibinfo {author} {\bibnamefont {{Gaudin, M.}}},\
  }\bibfield  {title} {\bibinfo {title} {Diagonalisation d'une classe
  d'hamiltoniens de spin},\ }\href
  {https://doi.org/10.1051/jphys:0197600370100108700} {\bibfield  {journal}
  {\bibinfo  {journal} {J. Phys. France}\ }\textbf {\bibinfo {volume} {37}},\
  \bibinfo {pages} {1087} (\bibinfo {year} {1976})}\BibitemShut {NoStop}%
\bibitem [{\citenamefont {Urbaszek}\ \emph {et~al.}(2013)\citenamefont
  {Urbaszek}, \citenamefont {Marie}, \citenamefont {Amand}, \citenamefont
  {Krebs}, \citenamefont {Voisin}, \citenamefont {Malentinsky}, \citenamefont
  {H\"ogele},\ and\ \citenamefont {Imamoglu}}]{urbaszek13}%
  \BibitemOpen
  \bibfield  {author} {\bibinfo {author} {\bibfnamefont {B.}~\bibnamefont
  {Urbaszek}}, \bibinfo {author} {\bibfnamefont {X.}~\bibnamefont {Marie}},
  \bibinfo {author} {\bibfnamefont {T.}~\bibnamefont {Amand}}, \bibinfo
  {author} {\bibfnamefont {O.}~\bibnamefont {Krebs}}, \bibinfo {author}
  {\bibfnamefont {P.}~\bibnamefont {Voisin}}, \bibinfo {author} {\bibfnamefont
  {P.}~\bibnamefont {Malentinsky}}, \bibinfo {author} {\bibfnamefont
  {A.}~\bibnamefont {H\"ogele}},\ and\ \bibinfo {author} {\bibfnamefont
  {A.}~\bibnamefont {Imamoglu}},\ }\bibfield  {title} {\bibinfo {title}
  {Nuclear spin physics in quantum dots: An optical investigation},\ }\href
  {https://doi.org/10.1103/RevModPhys.85.79} {\bibfield  {journal} {\bibinfo
  {journal} {Rev. Mod. Phys.}\ }\textbf {\bibinfo {volume} {85}},\ \bibinfo
  {pages} {79} (\bibinfo {year} {2013})}\BibitemShut {NoStop}%
\bibitem [{\citenamefont {Klauder}\ and\ \citenamefont
  {Anderson}(1962)}]{klauder62}%
  \BibitemOpen
  \bibfield  {author} {\bibinfo {author} {\bibfnamefont {J.~R.}\ \bibnamefont
  {Klauder}}\ and\ \bibinfo {author} {\bibfnamefont {P.~W.}\ \bibnamefont
  {Anderson}},\ }\bibfield  {title} {\bibinfo {title} {Spectral diffusion decay
  in spin resonance experiments},\ }\href
  {https://doi.org/10.1103/PhysRev.125.912} {\bibfield  {journal} {\bibinfo
  {journal} {Phys. Rev.}\ }\textbf {\bibinfo {volume} {125}},\ \bibinfo {pages}
  {912} (\bibinfo {year} {1962})}\BibitemShut {NoStop}%
\bibitem [{\citenamefont {de~Sousa}\ and\ \citenamefont
  {Das~Sarma}(2003)}]{desousa03}%
  \BibitemOpen
  \bibfield  {author} {\bibinfo {author} {\bibfnamefont {R.}~\bibnamefont
  {de~Sousa}}\ and\ \bibinfo {author} {\bibfnamefont {S.}~\bibnamefont
  {Das~Sarma}},\ }\bibfield  {title} {\bibinfo {title} {Theory of
  nuclear-induced spectral diffusion: Spin decoherence of phosphorus donors in
  si and gaas quantum dots},\ }\href
  {https://doi.org/10.1103/PhysRevB.68.115322} {\bibfield  {journal} {\bibinfo
  {journal} {Phys. Rev. B}\ }\textbf {\bibinfo {volume} {68}},\ \bibinfo
  {pages} {115322} (\bibinfo {year} {2003})}\BibitemShut {NoStop}%
\bibitem [{\citenamefont {Witzel}\ \emph {et~al.}(2005)\citenamefont {Witzel},
  \citenamefont {de~Sousa},\ and\ \citenamefont {Das~Sarma}}]{witzel05}%
  \BibitemOpen
  \bibfield  {author} {\bibinfo {author} {\bibfnamefont {W.~M.}\ \bibnamefont
  {Witzel}}, \bibinfo {author} {\bibfnamefont {R.}~\bibnamefont {de~Sousa}},\
  and\ \bibinfo {author} {\bibfnamefont {S.}~\bibnamefont {Das~Sarma}},\
  }\bibfield  {title} {\bibinfo {title} {Quantum theory of
  spectral-diffusion-induced electron spin decoherence},\ }\href
  {https://doi.org/10.1103/PhysRevB.72.161306} {\bibfield  {journal} {\bibinfo
  {journal} {Phys. Rev. B}\ }\textbf {\bibinfo {volume} {72}},\ \bibinfo
  {pages} {161306(R)} (\bibinfo {year} {2005})}\BibitemShut {NoStop}%
\bibitem [{\citenamefont {Witzel}\ and\ \citenamefont
  {Das~Sarma}(2006)}]{witzel06}%
  \BibitemOpen
  \bibfield  {author} {\bibinfo {author} {\bibfnamefont {W.~M.}\ \bibnamefont
  {Witzel}}\ and\ \bibinfo {author} {\bibfnamefont {S.}~\bibnamefont
  {Das~Sarma}},\ }\bibfield  {title} {\bibinfo {title} {Quantum theory for
  electron spin decoherence induced by nuclear spin dynamics in semiconductor
  quantum computer architectures: Spectral diffusion of localized electron
  spins in the nuclear solid-state environment},\ }\href
  {https://doi.org/10.1103/PhysRevB.74.035322} {\bibfield  {journal} {\bibinfo
  {journal} {Phys. Rev. B}\ }\textbf {\bibinfo {volume} {74}},\ \bibinfo
  {pages} {035322} (\bibinfo {year} {2006})}\BibitemShut {NoStop}%
\bibitem [{\citenamefont {Coish}\ and\ \citenamefont {Baugh}(2009)}]{coish09}%
  \BibitemOpen
  \bibfield  {author} {\bibinfo {author} {\bibfnamefont {W.~A.}\ \bibnamefont
  {Coish}}\ and\ \bibinfo {author} {\bibfnamefont {J.}~\bibnamefont {Baugh}},\
  }\bibfield  {title} {\bibinfo {title} {Nuclear spins in nanostructures},\
  }\href {https://doi.org/https://doi.org/10.1002/pssb.200945229} {\bibfield
  {journal} {\bibinfo  {journal} {physica status solidi (b)}\ }\textbf
  {\bibinfo {volume} {246}},\ \bibinfo {pages} {2203} (\bibinfo {year}
  {2009})},\ \Eprint
  {https://arxiv.org/abs/https://onlinelibrary.wiley.com/doi/pdf/10.1002/pssb.200945229}
  {https://onlinelibrary.wiley.com/doi/pdf/10.1002/pssb.200945229} \BibitemShut
  {NoStop}%
\bibitem [{\citenamefont {Cywi\ifmmode~\acute{n}\else
  \'{n}\fi{}ski}(2011)}]{cywinski11}%
  \BibitemOpen
  \bibfield  {author} {\bibinfo {author} {\bibfnamefont {L.}~\bibnamefont
  {Cywi\ifmmode~\acute{n}\else \'{n}\fi{}ski}},\ }\bibfield  {title} {\bibinfo
  {title} {Dephasing of electron spin qubits due to their interaction with
  nuclei in quantum dots},\ }\href {http://doi.org/10.12693/aphyspola.119.576}
  {\bibfield  {journal} {\bibinfo  {journal} {Acta Physica Polonica A}\
  }\textbf {\bibinfo {volume} {119}},\ \bibinfo {pages} {576–587} (\bibinfo
  {year} {2011})}\BibitemShut {NoStop}%
\bibitem [{\citenamefont {Tsyplyatyev}\ and\ \citenamefont
  {Loss}(2011)}]{oleksandr11}%
  \BibitemOpen
  \bibfield  {author} {\bibinfo {author} {\bibfnamefont {O.}~\bibnamefont
  {Tsyplyatyev}}\ and\ \bibinfo {author} {\bibfnamefont {D.}~\bibnamefont
  {Loss}},\ }\bibfield  {title} {\bibinfo {title} {Spectrum of an electron spin
  coupled to an unpolarized bath of nuclear spins},\ }\href
  {https://doi.org/10.1103/PhysRevLett.106.106803} {\bibfield  {journal}
  {\bibinfo  {journal} {Phys. Rev. Lett.}\ }\textbf {\bibinfo {volume} {106}},\
  \bibinfo {pages} {106803} (\bibinfo {year} {2011})}\BibitemShut {NoStop}%
\bibitem [{\citenamefont {Sinitsyn}\ \emph {et~al.}(2012)\citenamefont
  {Sinitsyn}, \citenamefont {Li}, \citenamefont {Crooker}, \citenamefont
  {Saxena},\ and\ \citenamefont {Smith}}]{sinitsyn12}%
  \BibitemOpen
  \bibfield  {author} {\bibinfo {author} {\bibfnamefont {N.~A.}\ \bibnamefont
  {Sinitsyn}}, \bibinfo {author} {\bibfnamefont {Y.}~\bibnamefont {Li}},
  \bibinfo {author} {\bibfnamefont {S.~A.}\ \bibnamefont {Crooker}}, \bibinfo
  {author} {\bibfnamefont {A.}~\bibnamefont {Saxena}},\ and\ \bibinfo {author}
  {\bibfnamefont {D.~L.}\ \bibnamefont {Smith}},\ }\bibfield  {title} {\bibinfo
  {title} {Role of nuclear quadrupole coupling on decoherence and relaxation of
  central spins in quantum dots},\ }\href
  {https://doi.org/10.1103/PhysRevLett.109.166605} {\bibfield  {journal}
  {\bibinfo  {journal} {Phys. Rev. Lett.}\ }\textbf {\bibinfo {volume} {109}},\
  \bibinfo {pages} {166605} (\bibinfo {year} {2012})}\BibitemShut {NoStop}%
\bibitem [{\citenamefont {Hahn}(1950)}]{hahn50}%
  \BibitemOpen
  \bibfield  {author} {\bibinfo {author} {\bibfnamefont {E.~L.}\ \bibnamefont
  {Hahn}},\ }\bibfield  {title} {\bibinfo {title} {Spin echoes},\ }\href
  {https://doi.org/10.1103/PhysRev.80.580} {\bibfield  {journal} {\bibinfo
  {journal} {Phys. Rev.}\ }\textbf {\bibinfo {volume} {80}},\ \bibinfo {pages}
  {580} (\bibinfo {year} {1950})}\BibitemShut {NoStop}%
\bibitem [{\citenamefont {Viola}\ and\ \citenamefont {Lloyd}(1998)}]{viola98}%
  \BibitemOpen
  \bibfield  {author} {\bibinfo {author} {\bibfnamefont {L.}~\bibnamefont
  {Viola}}\ and\ \bibinfo {author} {\bibfnamefont {S.}~\bibnamefont {Lloyd}},\
  }\bibfield  {title} {\bibinfo {title} {Dynamical suppression of decoherence
  in two-state quantum systems},\ }\href
  {https://doi.org/10.1103/PhysRevA.58.2733} {\bibfield  {journal} {\bibinfo
  {journal} {Phys. Rev. A}\ }\textbf {\bibinfo {volume} {58}},\ \bibinfo
  {pages} {2733} (\bibinfo {year} {1998})}\BibitemShut {NoStop}%
\bibitem [{\citenamefont {Carr}\ and\ \citenamefont {Purcell}(1954)}]{carr54}%
  \BibitemOpen
  \bibfield  {author} {\bibinfo {author} {\bibfnamefont {H.~Y.}\ \bibnamefont
  {Carr}}\ and\ \bibinfo {author} {\bibfnamefont {E.~M.}\ \bibnamefont
  {Purcell}},\ }\bibfield  {title} {\bibinfo {title} {Effects of diffusion on
  free precession in nuclear magnetic resonance experiments},\ }\href
  {https://doi.org/10.1103/PhysRev.94.630} {\bibfield  {journal} {\bibinfo
  {journal} {Phys. Rev.}\ }\textbf {\bibinfo {volume} {94}},\ \bibinfo {pages}
  {630} (\bibinfo {year} {1954})}\BibitemShut {NoStop}%
\bibitem [{\citenamefont {Uhrig}(2007)}]{uhrig07}%
  \BibitemOpen
  \bibfield  {author} {\bibinfo {author} {\bibfnamefont {G.~S.}\ \bibnamefont
  {Uhrig}},\ }\bibfield  {title} {\bibinfo {title} {Keeping a quantum bit alive
  by optimized $\ensuremath{\pi}$-pulse sequences},\ }\href
  {https://doi.org/10.1103/PhysRevLett.98.100504} {\bibfield  {journal}
  {\bibinfo  {journal} {Phys. Rev. Lett.}\ }\textbf {\bibinfo {volume} {98}},\
  \bibinfo {pages} {100504} (\bibinfo {year} {2007})}\BibitemShut {NoStop}%
\bibitem [{\citenamefont {Chekhovich}\ \emph {et~al.}(2015)\citenamefont
  {Chekhovich}, \citenamefont {Hopkinson}, \citenamefont {Skolnick},\ and\
  \citenamefont {Tartakovskii}}]{chekhovich15}%
  \BibitemOpen
  \bibfield  {author} {\bibinfo {author} {\bibfnamefont {E.~A.}\ \bibnamefont
  {Chekhovich}}, \bibinfo {author} {\bibfnamefont {M.}~\bibnamefont
  {Hopkinson}}, \bibinfo {author} {\bibfnamefont {M.~S.}\ \bibnamefont
  {Skolnick}},\ and\ \bibinfo {author} {\bibfnamefont {A.~I.}\ \bibnamefont
  {Tartakovskii}},\ }\bibfield  {title} {\bibinfo {title} {Suppression of
  nuclear spin bath fluctuations in self-assembled quantum dots induced by
  inhomogeneous strain},\ }\href {https://doi.org/10.1038/ncomms7348}
  {\bibfield  {journal} {\bibinfo  {journal} {Nature Communications}\ }\textbf
  {\bibinfo {volume} {6}},\ \bibinfo {pages} {6348} (\bibinfo {year}
  {2015})}\BibitemShut {NoStop}%
\bibitem [{\citenamefont {Cywi\ifmmode~\acute{n}\else \'{n}\fi{}ski}\ \emph
  {et~al.}(2008)\citenamefont {Cywi\ifmmode~\acute{n}\else \'{n}\fi{}ski},
  \citenamefont {Lutchyn}, \citenamefont {Nave},\ and\ \citenamefont
  {Das~Sarma}}]{cywinski08}%
  \BibitemOpen
  \bibfield  {author} {\bibinfo {author} {\bibfnamefont {L.}~\bibnamefont
  {Cywi\ifmmode~\acute{n}\else \'{n}\fi{}ski}}, \bibinfo {author}
  {\bibfnamefont {R.~M.}\ \bibnamefont {Lutchyn}}, \bibinfo {author}
  {\bibfnamefont {C.~P.}\ \bibnamefont {Nave}},\ and\ \bibinfo {author}
  {\bibfnamefont {S.}~\bibnamefont {Das~Sarma}},\ }\bibfield  {title} {\bibinfo
  {title} {How to enhance dephasing time in superconducting qubits},\ }\href
  {https://doi.org/10.1103/PhysRevB.77.174509} {\bibfield  {journal} {\bibinfo
  {journal} {Phys. Rev. B}\ }\textbf {\bibinfo {volume} {77}},\ \bibinfo
  {pages} {174509} (\bibinfo {year} {2008})}\BibitemShut {NoStop}%
\bibitem [{\citenamefont {Imamoglu}\ \emph {et~al.}(2003)\citenamefont
  {Imamoglu}, \citenamefont {Knill}, \citenamefont {Tian},\ and\ \citenamefont
  {Zoller}}]{imamoglu03}%
  \BibitemOpen
  \bibfield  {author} {\bibinfo {author} {\bibfnamefont {A.}~\bibnamefont
  {Imamoglu}}, \bibinfo {author} {\bibfnamefont {E.}~\bibnamefont {Knill}},
  \bibinfo {author} {\bibfnamefont {L.}~\bibnamefont {Tian}},\ and\ \bibinfo
  {author} {\bibfnamefont {P.}~\bibnamefont {Zoller}},\ }\bibfield  {title}
  {\bibinfo {title} {Optical pumping of quantum-dot nuclear spins},\ }\href
  {https://doi.org/10.1103/PhysRevLett.91.017402} {\bibfield  {journal}
  {\bibinfo  {journal} {Phys. Rev. Lett.}\ }\textbf {\bibinfo {volume} {91}},\
  \bibinfo {pages} {017402} (\bibinfo {year} {2003})}\BibitemShut {NoStop}%
\bibitem [{\citenamefont {H\"ogele}\ \emph {et~al.}(2012)\citenamefont
  {H\"ogele}, \citenamefont {Kroner}, \citenamefont {Latta}, \citenamefont
  {Claassen}, \citenamefont {Carusotto}, \citenamefont {Bulutay},\ and\
  \citenamefont {Imamoglu}}]{hogele12}%
  \BibitemOpen
  \bibfield  {author} {\bibinfo {author} {\bibfnamefont {A.}~\bibnamefont
  {H\"ogele}}, \bibinfo {author} {\bibfnamefont {M.}~\bibnamefont {Kroner}},
  \bibinfo {author} {\bibfnamefont {C.}~\bibnamefont {Latta}}, \bibinfo
  {author} {\bibfnamefont {M.}~\bibnamefont {Claassen}}, \bibinfo {author}
  {\bibfnamefont {I.}~\bibnamefont {Carusotto}}, \bibinfo {author}
  {\bibfnamefont {C.}~\bibnamefont {Bulutay}},\ and\ \bibinfo {author}
  {\bibfnamefont {A.}~\bibnamefont {Imamoglu}},\ }\bibfield  {title} {\bibinfo
  {title} {Dynamic nuclear spin polarization in the resonant laser excitation
  of an ingaas quantum dot},\ }\href
  {https://doi.org/10.1103/PhysRevLett.108.197403} {\bibfield  {journal}
  {\bibinfo  {journal} {Phys. Rev. Lett.}\ }\textbf {\bibinfo {volume} {108}},\
  \bibinfo {pages} {197403} (\bibinfo {year} {2012})}\BibitemShut {NoStop}%
\bibitem [{\citenamefont {Jacques}\ \emph {et~al.}(2009)\citenamefont
  {Jacques}, \citenamefont {Neumann}, \citenamefont {Beck}, \citenamefont
  {Markham}, \citenamefont {Twitchen}, \citenamefont {Meijer}, \citenamefont
  {Kaiser}, \citenamefont {Balasubramanian}, \citenamefont {Jelezko},\ and\
  \citenamefont {Wrachtrup}}]{jacques09}%
  \BibitemOpen
  \bibfield  {author} {\bibinfo {author} {\bibfnamefont {V.}~\bibnamefont
  {Jacques}}, \bibinfo {author} {\bibfnamefont {P.}~\bibnamefont {Neumann}},
  \bibinfo {author} {\bibfnamefont {J.}~\bibnamefont {Beck}}, \bibinfo {author}
  {\bibfnamefont {M.}~\bibnamefont {Markham}}, \bibinfo {author} {\bibfnamefont
  {D.}~\bibnamefont {Twitchen}}, \bibinfo {author} {\bibfnamefont
  {J.}~\bibnamefont {Meijer}}, \bibinfo {author} {\bibfnamefont
  {F.}~\bibnamefont {Kaiser}}, \bibinfo {author} {\bibfnamefont
  {G.}~\bibnamefont {Balasubramanian}}, \bibinfo {author} {\bibfnamefont
  {F.}~\bibnamefont {Jelezko}},\ and\ \bibinfo {author} {\bibfnamefont
  {J.}~\bibnamefont {Wrachtrup}},\ }\bibfield  {title} {\bibinfo {title}
  {Dynamic polarization of single nuclear spins by optical pumping of
  nitrogen-vacancy color centers in diamond at room temperature},\ }\href
  {https://doi.org/10.1103/PhysRevLett.102.057403} {\bibfield  {journal}
  {\bibinfo  {journal} {Phys. Rev. Lett.}\ }\textbf {\bibinfo {volume} {102}},\
  \bibinfo {pages} {057403} (\bibinfo {year} {2009})}\BibitemShut {NoStop}%
\bibitem [{\citenamefont {Falk}\ \emph {et~al.}(2015)\citenamefont {Falk},
  \citenamefont {Klimov}, \citenamefont {Iv\'ady}, \citenamefont {Sz\'asz},
  \citenamefont {Christle}, \citenamefont {Koehl}, \citenamefont {Gali},\ and\
  \citenamefont {Awschalom}}]{falk15}%
  \BibitemOpen
  \bibfield  {author} {\bibinfo {author} {\bibfnamefont {A.~L.}\ \bibnamefont
  {Falk}}, \bibinfo {author} {\bibfnamefont {P.~V.}\ \bibnamefont {Klimov}},
  \bibinfo {author} {\bibfnamefont {V.}~\bibnamefont {Iv\'ady}}, \bibinfo
  {author} {\bibfnamefont {K.}~\bibnamefont {Sz\'asz}}, \bibinfo {author}
  {\bibfnamefont {D.~J.}\ \bibnamefont {Christle}}, \bibinfo {author}
  {\bibfnamefont {W.~F.}\ \bibnamefont {Koehl}}, \bibinfo {author}
  {\bibfnamefont {A.}~\bibnamefont {Gali}},\ and\ \bibinfo {author}
  {\bibfnamefont {D.~D.}\ \bibnamefont {Awschalom}},\ }\bibfield  {title}
  {\bibinfo {title} {Optical polarization of nuclear spins in silicon
  carbide},\ }\href {https://doi.org/10.1103/PhysRevLett.114.247603} {\bibfield
   {journal} {\bibinfo  {journal} {Phys. Rev. Lett.}\ }\textbf {\bibinfo
  {volume} {114}},\ \bibinfo {pages} {247603} (\bibinfo {year}
  {2015})}\BibitemShut {NoStop}%
\bibitem [{\citenamefont {Ajoy}\ \emph {et~al.}(2019)\citenamefont {Ajoy},
  \citenamefont {Bissbort}, \citenamefont {Poletti},\ and\ \citenamefont
  {Cappellaro}}]{ajoy19}%
  \BibitemOpen
  \bibfield  {author} {\bibinfo {author} {\bibfnamefont {A.}~\bibnamefont
  {Ajoy}}, \bibinfo {author} {\bibfnamefont {U.}~\bibnamefont {Bissbort}},
  \bibinfo {author} {\bibfnamefont {D.}~\bibnamefont {Poletti}},\ and\ \bibinfo
  {author} {\bibfnamefont {P.}~\bibnamefont {Cappellaro}},\ }\bibfield  {title}
  {\bibinfo {title} {Selective decoupling and hamiltonian engineering in
  dipolar spin networks},\ }\href
  {https://doi.org/10.1103/PhysRevLett.122.013205} {\bibfield  {journal}
  {\bibinfo  {journal} {Phys. Rev. Lett.}\ }\textbf {\bibinfo {volume} {122}},\
  \bibinfo {pages} {013205} (\bibinfo {year} {2019})}\BibitemShut {NoStop}%
\bibitem [{\citenamefont {Taylor}\ \emph
  {et~al.}(2003{\natexlab{a}})\citenamefont {Taylor}, \citenamefont {Marcus},\
  and\ \citenamefont {Lukin}}]{taylor03-2}%
  \BibitemOpen
  \bibfield  {author} {\bibinfo {author} {\bibfnamefont {J.~M.}\ \bibnamefont
  {Taylor}}, \bibinfo {author} {\bibfnamefont {C.~M.}\ \bibnamefont {Marcus}},\
  and\ \bibinfo {author} {\bibfnamefont {M.~D.}\ \bibnamefont {Lukin}},\
  }\bibfield  {title} {\bibinfo {title} {Long-lived memory for mesoscopic
  quantum bits},\ }\href {https://doi.org/10.1103/PhysRevLett.90.206803}
  {\bibfield  {journal} {\bibinfo  {journal} {Phys. Rev. Lett.}\ }\textbf
  {\bibinfo {volume} {90}},\ \bibinfo {pages} {206803} (\bibinfo {year}
  {2003}{\natexlab{a}})}\BibitemShut {NoStop}%
\bibitem [{\citenamefont {Taylor}\ \emph
  {et~al.}(2003{\natexlab{b}})\citenamefont {Taylor}, \citenamefont
  {Imamoglu},\ and\ \citenamefont {Lukin}}]{taylor03}%
  \BibitemOpen
  \bibfield  {author} {\bibinfo {author} {\bibfnamefont {J.~M.}\ \bibnamefont
  {Taylor}}, \bibinfo {author} {\bibfnamefont {A.}~\bibnamefont {Imamoglu}},\
  and\ \bibinfo {author} {\bibfnamefont {M.~D.}\ \bibnamefont {Lukin}},\
  }\bibfield  {title} {\bibinfo {title} {Controlling a mesoscopic spin
  environment by quantum bit manipulation},\ }\href
  {https://doi.org/10.1103/PhysRevLett.91.246802} {\bibfield  {journal}
  {\bibinfo  {journal} {Phys. Rev. Lett.}\ }\textbf {\bibinfo {volume} {91}},\
  \bibinfo {pages} {246802} (\bibinfo {year} {2003}{\natexlab{b}})}\BibitemShut
  {NoStop}%
\bibitem [{\citenamefont {Gangloff}\ \emph {et~al.}(2019)\citenamefont
  {Gangloff}, \citenamefont {{\'E}thier-Majcher}, \citenamefont {Lang},
  \citenamefont {Denning}, \citenamefont {Bodey}, \citenamefont {Jackson},
  \citenamefont {Clarke}, \citenamefont {Hugues}, \citenamefont {Le~Gall},\
  and\ \citenamefont {Atat{\"u}re}}]{gangloff19}%
  \BibitemOpen
  \bibfield  {author} {\bibinfo {author} {\bibfnamefont {D.~A.}\ \bibnamefont
  {Gangloff}}, \bibinfo {author} {\bibfnamefont {G.}~\bibnamefont
  {{\'E}thier-Majcher}}, \bibinfo {author} {\bibfnamefont {C.}~\bibnamefont
  {Lang}}, \bibinfo {author} {\bibfnamefont {E.~V.}\ \bibnamefont {Denning}},
  \bibinfo {author} {\bibfnamefont {J.~H.}\ \bibnamefont {Bodey}}, \bibinfo
  {author} {\bibfnamefont {D.~M.}\ \bibnamefont {Jackson}}, \bibinfo {author}
  {\bibfnamefont {E.}~\bibnamefont {Clarke}}, \bibinfo {author} {\bibfnamefont
  {M.}~\bibnamefont {Hugues}}, \bibinfo {author} {\bibfnamefont
  {C.}~\bibnamefont {Le~Gall}},\ and\ \bibinfo {author} {\bibfnamefont
  {M.}~\bibnamefont {Atat{\"u}re}},\ }\bibfield  {title} {\bibinfo {title}
  {Quantum interface of an electron and a nuclear ensemble},\ }\href
  {https://doi.org/10.1126/science.aaw2906} {\bibfield  {journal} {\bibinfo
  {journal} {Science}\ }\textbf {\bibinfo {volume} {364}},\ \bibinfo {pages}
  {62} (\bibinfo {year} {2019})},\ \Eprint
  {https://arxiv.org/abs/https://science.sciencemag.org/content/364/6435/62.full.pdf}
  {https://science.sciencemag.org/content/364/6435/62.full.pdf} \BibitemShut
  {NoStop}%
\bibitem [{\citenamefont {Denning}\ \emph {et~al.}(2019)\citenamefont
  {Denning}, \citenamefont {Gangloff}, \citenamefont {Atat\"ure}, \citenamefont
  {M\o{}rk},\ and\ \citenamefont {Le~Gall}}]{denning19}%
  \BibitemOpen
  \bibfield  {author} {\bibinfo {author} {\bibfnamefont {E.~V.}\ \bibnamefont
  {Denning}}, \bibinfo {author} {\bibfnamefont {D.~A.}\ \bibnamefont
  {Gangloff}}, \bibinfo {author} {\bibfnamefont {M.}~\bibnamefont {Atat\"ure}},
  \bibinfo {author} {\bibfnamefont {J.}~\bibnamefont {M\o{}rk}},\ and\ \bibinfo
  {author} {\bibfnamefont {C.}~\bibnamefont {Le~Gall}},\ }\bibfield  {title}
  {\bibinfo {title} {Collective quantum memory activated by a driven central
  spin},\ }\href {https://doi.org/10.1103/PhysRevLett.123.140502} {\bibfield
  {journal} {\bibinfo  {journal} {Phys. Rev. Lett.}\ }\textbf {\bibinfo
  {volume} {123}},\ \bibinfo {pages} {140502} (\bibinfo {year}
  {2019})}\BibitemShut {NoStop}%
\bibitem [{\citenamefont {Dutt}\ \emph {et~al.}(2007)\citenamefont {Dutt},
  \citenamefont {Childress}, \citenamefont {Jiang}, \citenamefont {Togan},
  \citenamefont {Maze}, \citenamefont {Jelezko}, \citenamefont {Zibrov},
  \citenamefont {Hemmer},\ and\ \citenamefont {Lukin}}]{dutt07}%
  \BibitemOpen
  \bibfield  {author} {\bibinfo {author} {\bibfnamefont {M.~V.~G.}\
  \bibnamefont {Dutt}}, \bibinfo {author} {\bibfnamefont {L.}~\bibnamefont
  {Childress}}, \bibinfo {author} {\bibfnamefont {L.}~\bibnamefont {Jiang}},
  \bibinfo {author} {\bibfnamefont {E.}~\bibnamefont {Togan}}, \bibinfo
  {author} {\bibfnamefont {J.}~\bibnamefont {Maze}}, \bibinfo {author}
  {\bibfnamefont {F.}~\bibnamefont {Jelezko}}, \bibinfo {author} {\bibfnamefont
  {A.~S.}\ \bibnamefont {Zibrov}}, \bibinfo {author} {\bibfnamefont {P.~R.}\
  \bibnamefont {Hemmer}},\ and\ \bibinfo {author} {\bibfnamefont {M.~D.}\
  \bibnamefont {Lukin}},\ }\bibfield  {title} {\bibinfo {title} {Quantum
  register based on individual electronic and nuclear spin qubits in diamond},\
  }\href {https://doi.org/10.1126/science.1139831} {\bibfield  {journal}
  {\bibinfo  {journal} {Science}\ }\textbf {\bibinfo {volume} {316}},\ \bibinfo
  {pages} {1312} (\bibinfo {year} {2007})},\ \Eprint
  {https://arxiv.org/abs/https://science.sciencemag.org/content/316/5829/1312.full.pdf}
  {https://science.sciencemag.org/content/316/5829/1312.full.pdf} \BibitemShut
  {NoStop}%
\bibitem [{\citenamefont {Neumann}\ \emph {et~al.}(2008)\citenamefont
  {Neumann}, \citenamefont {Mizuochi}, \citenamefont {Rempp}, \citenamefont
  {Hemmer}, \citenamefont {Watanabe}, \citenamefont {Yamasaki}, \citenamefont
  {Jacques}, \citenamefont {Gaebel}, \citenamefont {Jelezko},\ and\
  \citenamefont {Wrachtrup}}]{neuman08}%
  \BibitemOpen
  \bibfield  {author} {\bibinfo {author} {\bibfnamefont {P.}~\bibnamefont
  {Neumann}}, \bibinfo {author} {\bibfnamefont {N.}~\bibnamefont {Mizuochi}},
  \bibinfo {author} {\bibfnamefont {F.}~\bibnamefont {Rempp}}, \bibinfo
  {author} {\bibfnamefont {P.}~\bibnamefont {Hemmer}}, \bibinfo {author}
  {\bibfnamefont {H.}~\bibnamefont {Watanabe}}, \bibinfo {author}
  {\bibfnamefont {S.}~\bibnamefont {Yamasaki}}, \bibinfo {author}
  {\bibfnamefont {V.}~\bibnamefont {Jacques}}, \bibinfo {author} {\bibfnamefont
  {T.}~\bibnamefont {Gaebel}}, \bibinfo {author} {\bibfnamefont
  {F.}~\bibnamefont {Jelezko}},\ and\ \bibinfo {author} {\bibfnamefont
  {J.}~\bibnamefont {Wrachtrup}},\ }\bibfield  {title} {\bibinfo {title}
  {Multipartite entanglement among single spins in diamond},\ }\href
  {https://doi.org/10.1126/science.1157233} {\bibfield  {journal} {\bibinfo
  {journal} {Science}\ }\textbf {\bibinfo {volume} {320}},\ \bibinfo {pages}
  {1326} (\bibinfo {year} {2008})},\ \Eprint
  {https://arxiv.org/abs/https://science.sciencemag.org/content/320/5881/1326.full.pdf}
  {https://science.sciencemag.org/content/320/5881/1326.full.pdf} \BibitemShut
  {NoStop}%
\bibitem [{\citenamefont {Fuchs}\ \emph {et~al.}(2011)\citenamefont {Fuchs},
  \citenamefont {Burkard}, \citenamefont {Klimov},\ and\ \citenamefont
  {Awschalom}}]{fuchs11}%
  \BibitemOpen
  \bibfield  {author} {\bibinfo {author} {\bibfnamefont {G.~D.}\ \bibnamefont
  {Fuchs}}, \bibinfo {author} {\bibfnamefont {G.}~\bibnamefont {Burkard}},
  \bibinfo {author} {\bibfnamefont {P.~V.}\ \bibnamefont {Klimov}},\ and\
  \bibinfo {author} {\bibfnamefont {D.~D.}\ \bibnamefont {Awschalom}},\
  }\bibfield  {title} {\bibinfo {title} {A quantum memory intrinsic to single
  nitrogen--vacancy centres in diamond},\ }\href
  {https://doi.org/10.1038/nphys2026} {\bibfield  {journal} {\bibinfo
  {journal} {Nature Physics}\ }\textbf {\bibinfo {volume} {7}},\ \bibinfo
  {pages} {789} (\bibinfo {year} {2011})}\BibitemShut {NoStop}%
\bibitem [{\citenamefont {Taminiau}\ \emph {et~al.}(2014)\citenamefont
  {Taminiau}, \citenamefont {Cramer}, \citenamefont {van~der Sar},
  \citenamefont {Dobrovitski},\ and\ \citenamefont {Hanson}}]{taminiau14}%
  \BibitemOpen
  \bibfield  {author} {\bibinfo {author} {\bibfnamefont {T.~H.}\ \bibnamefont
  {Taminiau}}, \bibinfo {author} {\bibfnamefont {J.}~\bibnamefont {Cramer}},
  \bibinfo {author} {\bibfnamefont {T.}~\bibnamefont {van~der Sar}}, \bibinfo
  {author} {\bibfnamefont {V.~V.}\ \bibnamefont {Dobrovitski}},\ and\ \bibinfo
  {author} {\bibfnamefont {R.}~\bibnamefont {Hanson}},\ }\bibfield  {title}
  {\bibinfo {title} {Universal control and error correction in multi-qubit spin
  registers in diamond},\ }\href {https://doi.org/10.1038/nnano.2014.2}
  {\bibfield  {journal} {\bibinfo  {journal} {Nature Nanotechnology}\ }\textbf
  {\bibinfo {volume} {9}},\ \bibinfo {pages} {171} (\bibinfo {year}
  {2014})}\BibitemShut {NoStop}%
\bibitem [{\citenamefont {Bradley}\ \emph {et~al.}(2019)\citenamefont
  {Bradley}, \citenamefont {Randall}, \citenamefont {Abobeih}, \citenamefont
  {Berrevoets}, \citenamefont {Degen}, \citenamefont {Bakker}, \citenamefont
  {Markham}, \citenamefont {Twitchen},\ and\ \citenamefont
  {Taminiau}}]{bradley19}%
  \BibitemOpen
  \bibfield  {author} {\bibinfo {author} {\bibfnamefont {C.~E.}\ \bibnamefont
  {Bradley}}, \bibinfo {author} {\bibfnamefont {J.}~\bibnamefont {Randall}},
  \bibinfo {author} {\bibfnamefont {M.~H.}\ \bibnamefont {Abobeih}}, \bibinfo
  {author} {\bibfnamefont {R.~C.}\ \bibnamefont {Berrevoets}}, \bibinfo
  {author} {\bibfnamefont {M.~J.}\ \bibnamefont {Degen}}, \bibinfo {author}
  {\bibfnamefont {M.~A.}\ \bibnamefont {Bakker}}, \bibinfo {author}
  {\bibfnamefont {M.}~\bibnamefont {Markham}}, \bibinfo {author} {\bibfnamefont
  {D.~J.}\ \bibnamefont {Twitchen}},\ and\ \bibinfo {author} {\bibfnamefont
  {T.~H.}\ \bibnamefont {Taminiau}},\ }\bibfield  {title} {\bibinfo {title} {A
  ten-qubit solid-state spin register with quantum memory up to one minute},\
  }\href {https://doi.org/10.1103/PhysRevX.9.031045} {\bibfield  {journal}
  {\bibinfo  {journal} {Phys. Rev. X}\ }\textbf {\bibinfo {volume} {9}},\
  \bibinfo {pages} {031045} (\bibinfo {year} {2019})}\BibitemShut {NoStop}%
\bibitem [{\citenamefont {Bulutay}\ \emph {et~al.}(2014)\citenamefont
  {Bulutay}, \citenamefont {Chekhovich},\ and\ \citenamefont
  {Tartakovskii}}]{bulutay14}%
  \BibitemOpen
  \bibfield  {author} {\bibinfo {author} {\bibfnamefont {C.}~\bibnamefont
  {Bulutay}}, \bibinfo {author} {\bibfnamefont {E.~A.}\ \bibnamefont
  {Chekhovich}},\ and\ \bibinfo {author} {\bibfnamefont {A.~I.}\ \bibnamefont
  {Tartakovskii}},\ }\bibfield  {title} {\bibinfo {title} {Nuclear magnetic
  resonance inverse spectra of ingaas quantum dots: Atomistic level structural
  information},\ }\href {https://doi.org/10.1103/PhysRevB.90.205425} {\bibfield
   {journal} {\bibinfo  {journal} {Phys. Rev. B}\ }\textbf {\bibinfo {volume}
  {90}},\ \bibinfo {pages} {205425} (\bibinfo {year} {2014})}\BibitemShut
  {NoStop}%
\bibitem [{\citenamefont {Ajoy}\ \emph {et~al.}(2015)\citenamefont {Ajoy},
  \citenamefont {Bissbort}, \citenamefont {Lukin}, \citenamefont {Walsworth},\
  and\ \citenamefont {Cappellaro}}]{ajoy15}%
  \BibitemOpen
  \bibfield  {author} {\bibinfo {author} {\bibfnamefont {A.}~\bibnamefont
  {Ajoy}}, \bibinfo {author} {\bibfnamefont {U.}~\bibnamefont {Bissbort}},
  \bibinfo {author} {\bibfnamefont {M.~D.}\ \bibnamefont {Lukin}}, \bibinfo
  {author} {\bibfnamefont {R.~L.}\ \bibnamefont {Walsworth}},\ and\ \bibinfo
  {author} {\bibfnamefont {P.}~\bibnamefont {Cappellaro}},\ }\bibfield  {title}
  {\bibinfo {title} {Atomic-scale nuclear spin imaging using quantum-assisted
  sensors in diamond},\ }\href {https://doi.org/10.1103/PhysRevX.5.011001}
  {\bibfield  {journal} {\bibinfo  {journal} {Phys. Rev. X}\ }\textbf {\bibinfo
  {volume} {5}},\ \bibinfo {pages} {011001} (\bibinfo {year}
  {2015})}\BibitemShut {NoStop}%
\bibitem [{\citenamefont {Kost}\ \emph {et~al.}(2015)\citenamefont {Kost},
  \citenamefont {Cai},\ and\ \citenamefont {Plenio}}]{kost15}%
  \BibitemOpen
  \bibfield  {author} {\bibinfo {author} {\bibfnamefont {M.}~\bibnamefont
  {Kost}}, \bibinfo {author} {\bibfnamefont {J.}~\bibnamefont {Cai}},\ and\
  \bibinfo {author} {\bibfnamefont {M.~B.}\ \bibnamefont {Plenio}},\ }\bibfield
   {title} {\bibinfo {title} {Resolving single molecule structures with
  {Nitrogen}-vacancy centers in diamond},\ }\href
  {https://doi.org/10.1038/srep11007} {\bibfield  {journal} {\bibinfo
  {journal} {Scientific Reports}\ }\textbf {\bibinfo {volume} {5}},\ \bibinfo
  {pages} {11007} (\bibinfo {year} {2015})}\BibitemShut {NoStop}%
\bibitem [{\citenamefont {Wang}\ \emph {et~al.}(2016)\citenamefont {Wang},
  \citenamefont {Haase}, \citenamefont {Casanova},\ and\ \citenamefont
  {Plenio}}]{wang16}%
  \BibitemOpen
  \bibfield  {author} {\bibinfo {author} {\bibfnamefont {Z.-Y.}\ \bibnamefont
  {Wang}}, \bibinfo {author} {\bibfnamefont {J.~F.}\ \bibnamefont {Haase}},
  \bibinfo {author} {\bibfnamefont {J.}~\bibnamefont {Casanova}},\ and\
  \bibinfo {author} {\bibfnamefont {M.~B.}\ \bibnamefont {Plenio}},\ }\bibfield
   {title} {\bibinfo {title} {Positioning nuclear spins in interacting clusters
  for quantum technologies and bioimaging},\ }\href
  {https://doi.org/10.1103/PhysRevB.93.174104} {\bibfield  {journal} {\bibinfo
  {journal} {Phys. Rev. B}\ }\textbf {\bibinfo {volume} {93}},\ \bibinfo
  {pages} {174104} (\bibinfo {year} {2016})}\BibitemShut {NoStop}%
\bibitem [{\citenamefont {Perunicic}\ \emph {et~al.}(2016)\citenamefont
  {Perunicic}, \citenamefont {Hill}, \citenamefont {Hall},\ and\ \citenamefont
  {Hollenberg}}]{perunicic16}%
  \BibitemOpen
  \bibfield  {author} {\bibinfo {author} {\bibfnamefont {V.~S.}\ \bibnamefont
  {Perunicic}}, \bibinfo {author} {\bibfnamefont {C.~D.}\ \bibnamefont {Hill}},
  \bibinfo {author} {\bibfnamefont {L.~T.}\ \bibnamefont {Hall}},\ and\
  \bibinfo {author} {\bibfnamefont {L.~C.~L.}\ \bibnamefont {Hollenberg}},\
  }\bibfield  {title} {\bibinfo {title} {A quantum spin-probe molecular
  microscope},\ }\href {https://doi.org/10.1038/ncomms12667} {\bibfield
  {journal} {\bibinfo  {journal} {Nature Communications}\ }\textbf {\bibinfo
  {volume} {7}},\ \bibinfo {pages} {12667} (\bibinfo {year}
  {2016})}\BibitemShut {NoStop}%
\bibitem [{\citenamefont {Maze}\ \emph
  {et~al.}(2008{\natexlab{a}})\citenamefont {Maze}, \citenamefont {Stanwix},
  \citenamefont {Hodges}, \citenamefont {Hong}, \citenamefont {Taylor},
  \citenamefont {Cappellaro}, \citenamefont {Jiang}, \citenamefont {Dutt},
  \citenamefont {Togan}, \citenamefont {Zibrov} \emph {et~al.}}]{maze08a}%
  \BibitemOpen
  \bibfield  {author} {\bibinfo {author} {\bibfnamefont {J.~R.}\ \bibnamefont
  {Maze}}, \bibinfo {author} {\bibfnamefont {P.~L.}\ \bibnamefont {Stanwix}},
  \bibinfo {author} {\bibfnamefont {J.~S.}\ \bibnamefont {Hodges}}, \bibinfo
  {author} {\bibfnamefont {S.}~\bibnamefont {Hong}}, \bibinfo {author}
  {\bibfnamefont {J.~M.}\ \bibnamefont {Taylor}}, \bibinfo {author}
  {\bibfnamefont {P.}~\bibnamefont {Cappellaro}}, \bibinfo {author}
  {\bibfnamefont {L.}~\bibnamefont {Jiang}}, \bibinfo {author} {\bibfnamefont
  {M.~G.}\ \bibnamefont {Dutt}}, \bibinfo {author} {\bibfnamefont
  {E.}~\bibnamefont {Togan}}, \bibinfo {author} {\bibfnamefont
  {A.}~\bibnamefont {Zibrov}}, \emph {et~al.},\ }\bibfield  {title} {\bibinfo
  {title} {Nanoscale magnetic sensing with an individual electronic spin in
  diamond},\ }\href {https://doi.org/10.1038/nature07279} {\bibfield  {journal}
  {\bibinfo  {journal} {Nature}\ }\textbf {\bibinfo {volume} {455}},\ \bibinfo
  {pages} {644} (\bibinfo {year} {2008}{\natexlab{a}})}\BibitemShut {NoStop}%
\bibitem [{\citenamefont {Meriles}\ \emph {et~al.}(2010)\citenamefont
  {Meriles}, \citenamefont {Jiang}, \citenamefont {Goldstein}, \citenamefont
  {Hodges}, \citenamefont {Maze}, \citenamefont {Lukin},\ and\ \citenamefont
  {Cappellaro}}]{meriles10}%
  \BibitemOpen
  \bibfield  {author} {\bibinfo {author} {\bibfnamefont {C.~A.}\ \bibnamefont
  {Meriles}}, \bibinfo {author} {\bibfnamefont {L.}~\bibnamefont {Jiang}},
  \bibinfo {author} {\bibfnamefont {G.}~\bibnamefont {Goldstein}}, \bibinfo
  {author} {\bibfnamefont {J.~S.}\ \bibnamefont {Hodges}}, \bibinfo {author}
  {\bibfnamefont {J.}~\bibnamefont {Maze}}, \bibinfo {author} {\bibfnamefont
  {M.~D.}\ \bibnamefont {Lukin}},\ and\ \bibinfo {author} {\bibfnamefont
  {P.}~\bibnamefont {Cappellaro}},\ }\bibfield  {title} {\bibinfo {title}
  {Imaging mesoscopic nuclear spin noise with a diamond magnetometer},\ }\href
  {https://doi.org/10.1063/1.3483676} {\bibfield  {journal} {\bibinfo
  {journal} {The Journal of Chemical Physics}\ }\textbf {\bibinfo {volume}
  {133}},\ \bibinfo {pages} {124105} (\bibinfo {year} {2010})},\ \Eprint
  {https://arxiv.org/abs/https://doi.org/10.1063/1.3483676}
  {https://doi.org/10.1063/1.3483676} \BibitemShut {NoStop}%
\bibitem [{\citenamefont {Arai}\ \emph {et~al.}(2015)\citenamefont {Arai},
  \citenamefont {Belthangady}, \citenamefont {Zhang}, \citenamefont {Bar-Gill},
  \citenamefont {DeVience}, \citenamefont {Cappellaro}, \citenamefont
  {Yacoby},\ and\ \citenamefont {Walsworth}}]{arai15}%
  \BibitemOpen
  \bibfield  {author} {\bibinfo {author} {\bibfnamefont {K.}~\bibnamefont
  {Arai}}, \bibinfo {author} {\bibfnamefont {C.}~\bibnamefont {Belthangady}},
  \bibinfo {author} {\bibfnamefont {H.}~\bibnamefont {Zhang}}, \bibinfo
  {author} {\bibfnamefont {N.}~\bibnamefont {Bar-Gill}}, \bibinfo {author}
  {\bibfnamefont {S.}~\bibnamefont {DeVience}}, \bibinfo {author}
  {\bibfnamefont {P.}~\bibnamefont {Cappellaro}}, \bibinfo {author}
  {\bibfnamefont {A.}~\bibnamefont {Yacoby}},\ and\ \bibinfo {author}
  {\bibfnamefont {R.~L.}\ \bibnamefont {Walsworth}},\ }\bibfield  {title}
  {\bibinfo {title} {Fourier magnetic imaging with nanoscale resolution and
  compressed sensing speed-up using electronic spins in diamond},\ }\href
  {https://doi.org/10.1038/nnano.2015.171} {\bibfield  {journal} {\bibinfo
  {journal} {Nature nanotechnology}\ }\textbf {\bibinfo {volume} {10}},\
  \bibinfo {pages} {859} (\bibinfo {year} {2015})}\BibitemShut {NoStop}%
\bibitem [{\citenamefont {Abobeih}\ \emph {et~al.}(2019)\citenamefont
  {Abobeih}, \citenamefont {Randall}, \citenamefont {Bradley}, \citenamefont
  {Bartling}, \citenamefont {Bakker}, \citenamefont {Degen}, \citenamefont
  {Markham}, \citenamefont {Twitchen},\ and\ \citenamefont
  {Taminiau}}]{abobeih19}%
  \BibitemOpen
  \bibfield  {author} {\bibinfo {author} {\bibfnamefont {M.~H.}\ \bibnamefont
  {Abobeih}}, \bibinfo {author} {\bibfnamefont {J.}~\bibnamefont {Randall}},
  \bibinfo {author} {\bibfnamefont {C.~E.}\ \bibnamefont {Bradley}}, \bibinfo
  {author} {\bibfnamefont {H.~P.}\ \bibnamefont {Bartling}}, \bibinfo {author}
  {\bibfnamefont {M.~A.}\ \bibnamefont {Bakker}}, \bibinfo {author}
  {\bibfnamefont {M.~J.}\ \bibnamefont {Degen}}, \bibinfo {author}
  {\bibfnamefont {M.}~\bibnamefont {Markham}}, \bibinfo {author} {\bibfnamefont
  {D.~J.}\ \bibnamefont {Twitchen}},\ and\ \bibinfo {author} {\bibfnamefont
  {T.~H.}\ \bibnamefont {Taminiau}},\ }\bibfield  {title} {\bibinfo {title}
  {Atomic-scale imaging of a 27-nuclear-spin cluster using a quantum sensor},\
  }\href {https://doi.org/10.1038/s41586-019-1834-7} {\bibfield  {journal}
  {\bibinfo  {journal} {Nature}\ }\textbf {\bibinfo {volume} {576}},\ \bibinfo
  {pages} {411} (\bibinfo {year} {2019})}\BibitemShut {NoStop}%
\bibitem [{\citenamefont {Yao}\ \emph {et~al.}(2006)\citenamefont {Yao},
  \citenamefont {Liu},\ and\ \citenamefont {Sham}}]{yao06}%
  \BibitemOpen
  \bibfield  {author} {\bibinfo {author} {\bibfnamefont {W.}~\bibnamefont
  {Yao}}, \bibinfo {author} {\bibfnamefont {R.-B.}\ \bibnamefont {Liu}},\ and\
  \bibinfo {author} {\bibfnamefont {L.~J.}\ \bibnamefont {Sham}},\ }\bibfield
  {title} {\bibinfo {title} {Theory of electron spin decoherence by interacting
  nuclear spins in a quantum dot},\ }\href
  {https://doi.org/10.1103/PhysRevB.74.195301} {\bibfield  {journal} {\bibinfo
  {journal} {Phys. Rev. B}\ }\textbf {\bibinfo {volume} {74}},\ \bibinfo
  {pages} {195301} (\bibinfo {year} {2006})}\BibitemShut {NoStop}%
\bibitem [{\citenamefont {Liu}\ \emph {et~al.}(2007)\citenamefont {Liu},
  \citenamefont {Yao},\ and\ \citenamefont {Sham}}]{liu07}%
  \BibitemOpen
  \bibfield  {author} {\bibinfo {author} {\bibfnamefont {R.-B.}\ \bibnamefont
  {Liu}}, \bibinfo {author} {\bibfnamefont {W.}~\bibnamefont {Yao}},\ and\
  \bibinfo {author} {\bibfnamefont {L.~J.}\ \bibnamefont {Sham}},\ }\bibfield
  {title} {\bibinfo {title} {Control of electron spin decoherence caused by
  electron{\textendash}nuclear spin dynamics in a quantum dot},\ }\href
  {https://doi.org/10.1088/1367-2630/9/7/226} {\bibfield  {journal} {\bibinfo
  {journal} {New Journal of Physics}\ }\textbf {\bibinfo {volume} {9}},\
  \bibinfo {pages} {226} (\bibinfo {year} {2007})}\BibitemShut {NoStop}%
\bibitem [{\citenamefont {Yao}\ \emph {et~al.}(2007)\citenamefont {Yao},
  \citenamefont {Liu},\ and\ \citenamefont {Sham}}]{yao07}%
  \BibitemOpen
  \bibfield  {author} {\bibinfo {author} {\bibfnamefont {W.}~\bibnamefont
  {Yao}}, \bibinfo {author} {\bibfnamefont {R.-B.}\ \bibnamefont {Liu}},\ and\
  \bibinfo {author} {\bibfnamefont {L.~J.}\ \bibnamefont {Sham}},\ }\bibfield
  {title} {\bibinfo {title} {Restoring coherence lost to a slow interacting
  mesoscopic spin bath},\ }\href
  {https://doi.org/10.1103/PhysRevLett.98.077602} {\bibfield  {journal}
  {\bibinfo  {journal} {Phys. Rev. Lett.}\ }\textbf {\bibinfo {volume} {98}},\
  \bibinfo {pages} {077602} (\bibinfo {year} {2007})}\BibitemShut {NoStop}%
\bibitem [{\citenamefont {Saikin}\ \emph {et~al.}(2007)\citenamefont {Saikin},
  \citenamefont {Yao},\ and\ \citenamefont {Sham}}]{saikin07}%
  \BibitemOpen
  \bibfield  {author} {\bibinfo {author} {\bibfnamefont {S.~K.}\ \bibnamefont
  {Saikin}}, \bibinfo {author} {\bibfnamefont {W.}~\bibnamefont {Yao}},\ and\
  \bibinfo {author} {\bibfnamefont {L.~J.}\ \bibnamefont {Sham}},\ }\bibfield
  {title} {\bibinfo {title} {Single-electron spin decoherence by nuclear spin
  bath: Linked-cluster expansion approach},\ }\href
  {https://doi.org/10.1103/PhysRevB.75.125314} {\bibfield  {journal} {\bibinfo
  {journal} {Phys. Rev. B}\ }\textbf {\bibinfo {volume} {75}},\ \bibinfo
  {pages} {125314} (\bibinfo {year} {2007})}\BibitemShut {NoStop}%
\bibitem [{\citenamefont {Yang}\ and\ \citenamefont {Liu}(2008)}]{yang08}%
  \BibitemOpen
  \bibfield  {author} {\bibinfo {author} {\bibfnamefont {W.}~\bibnamefont
  {Yang}}\ and\ \bibinfo {author} {\bibfnamefont {R.-B.}\ \bibnamefont {Liu}},\
  }\bibfield  {title} {\bibinfo {title} {Quantum many-body theory of qubit
  decoherence in a finite-size spin bath},\ }\href
  {https://doi.org/10.1103/PhysRevB.78.085315} {\bibfield  {journal} {\bibinfo
  {journal} {Phys. Rev. B}\ }\textbf {\bibinfo {volume} {78}},\ \bibinfo
  {pages} {085315} (\bibinfo {year} {2008})}\BibitemShut {NoStop}%
\bibitem [{\citenamefont {Yang}\ and\ \citenamefont {Liu}(2009)}]{yang09}%
  \BibitemOpen
  \bibfield  {author} {\bibinfo {author} {\bibfnamefont {W.}~\bibnamefont
  {Yang}}\ and\ \bibinfo {author} {\bibfnamefont {R.-B.}\ \bibnamefont {Liu}},\
  }\bibfield  {title} {\bibinfo {title} {Quantum many-body theory of qubit
  decoherence in a finite-size spin bath. ii. ensemble dynamics},\ }\href
  {https://doi.org/10.1103/PhysRevB.79.115320} {\bibfield  {journal} {\bibinfo
  {journal} {Phys. Rev. B}\ }\textbf {\bibinfo {volume} {79}},\ \bibinfo
  {pages} {115320} (\bibinfo {year} {2009})}\BibitemShut {NoStop}%
\bibitem [{\citenamefont {Maze}\ \emph
  {et~al.}(2008{\natexlab{b}})\citenamefont {Maze}, \citenamefont {Taylor},\
  and\ \citenamefont {Lukin}}]{maze08b}%
  \BibitemOpen
  \bibfield  {author} {\bibinfo {author} {\bibfnamefont {J.~R.}\ \bibnamefont
  {Maze}}, \bibinfo {author} {\bibfnamefont {J.~M.}\ \bibnamefont {Taylor}},\
  and\ \bibinfo {author} {\bibfnamefont {M.~D.}\ \bibnamefont {Lukin}},\
  }\bibfield  {title} {\bibinfo {title} {Electron spin decoherence of single
  nitrogen-vacancy defects in diamond},\ }\href
  {https://doi.org/10.1103/PhysRevB.78.094303} {\bibfield  {journal} {\bibinfo
  {journal} {Phys. Rev. B}\ }\textbf {\bibinfo {volume} {78}},\ \bibinfo
  {pages} {094303} (\bibinfo {year} {2008}{\natexlab{b}})}\BibitemShut
  {NoStop}%
\bibitem [{\citenamefont {Hall}\ \emph {et~al.}(2014)\citenamefont {Hall},
  \citenamefont {Cole},\ and\ \citenamefont {Hollenberg}}]{hall14}%
  \BibitemOpen
  \bibfield  {author} {\bibinfo {author} {\bibfnamefont {L.~T.}\ \bibnamefont
  {Hall}}, \bibinfo {author} {\bibfnamefont {J.~H.}\ \bibnamefont {Cole}},\
  and\ \bibinfo {author} {\bibfnamefont {L.~C.~L.}\ \bibnamefont
  {Hollenberg}},\ }\bibfield  {title} {\bibinfo {title} {Analytic solutions to
  the central-spin problem for nitrogen-vacancy centers in diamond},\ }\href
  {https://doi.org/10.1103/PhysRevB.90.075201} {\bibfield  {journal} {\bibinfo
  {journal} {Phys. Rev. B}\ }\textbf {\bibinfo {volume} {90}},\ \bibinfo
  {pages} {075201} (\bibinfo {year} {2014})}\BibitemShut {NoStop}%
\bibitem [{\citenamefont {Cywi\ifmmode~\acute{n}\else \'{n}\fi{}ski}\ \emph
  {et~al.}(2009{\natexlab{b}})\citenamefont {Cywi\ifmmode~\acute{n}\else
  \'{n}\fi{}ski}, \citenamefont {Witzel},\ and\ \citenamefont
  {Das~Sarma}}]{cywinskiprb09}%
  \BibitemOpen
  \bibfield  {author} {\bibinfo {author} {\bibfnamefont {L.}~\bibnamefont
  {Cywi\ifmmode~\acute{n}\else \'{n}\fi{}ski}}, \bibinfo {author}
  {\bibfnamefont {W.~M.}\ \bibnamefont {Witzel}},\ and\ \bibinfo {author}
  {\bibfnamefont {S.}~\bibnamefont {Das~Sarma}},\ }\bibfield  {title} {\bibinfo
  {title} {Pure quantum dephasing of a solid-state electron spin qubit in a
  large nuclear spin bath coupled by long-range hyperfine-mediated
  interactions},\ }\href {https://doi.org/10.1103/PhysRevB.79.245314}
  {\bibfield  {journal} {\bibinfo  {journal} {Phys. Rev. B}\ }\textbf {\bibinfo
  {volume} {79}},\ \bibinfo {pages} {245314} (\bibinfo {year}
  {2009}{\natexlab{b}})}\BibitemShut {NoStop}%
\bibitem [{\citenamefont {Yang}\ \emph {et~al.}(2017)\citenamefont {Yang},
  \citenamefont {Ma},\ and\ \citenamefont {Liu}}]{yang17}%
  \BibitemOpen
  \bibfield  {author} {\bibinfo {author} {\bibfnamefont {W.}~\bibnamefont
  {Yang}}, \bibinfo {author} {\bibfnamefont {W.-L.}\ \bibnamefont {Ma}},\ and\
  \bibinfo {author} {\bibfnamefont {R.-B.}\ \bibnamefont {Liu}},\ }\bibfield
  {title} {\bibinfo {title} {Quantum many-body theory for electron spin
  decoherence in nanoscale nuclear spin baths},\ }\href
  {https://doi.org/10.1088/0034-4885/80/1/016001} {\bibfield  {journal}
  {\bibinfo  {journal} {Reports on Progress in Physics}\ }\textbf {\bibinfo
  {volume} {80}},\ \bibinfo {pages} {016001} (\bibinfo {year}
  {2017})}\BibitemShut {NoStop}%
\bibitem [{\citenamefont {Kwiatkowski}\ and\ \citenamefont
  {Cywi\ifmmode~\acute{n}\else \'{n}\fi{}ski}(2018)}]{kwiatkowski18}%
  \BibitemOpen
  \bibfield  {author} {\bibinfo {author} {\bibfnamefont {D.}~\bibnamefont
  {Kwiatkowski}}\ and\ \bibinfo {author} {\bibfnamefont {L.}~\bibnamefont
  {Cywi\ifmmode~\acute{n}\else \'{n}\fi{}ski}},\ }\bibfield  {title} {\bibinfo
  {title} {Decoherence of two entangled spin qubits coupled to an interacting
  sparse nuclear spin bath: Application to nitrogen vacancy centers},\ }\href
  {https://doi.org/10.1103/PhysRevB.98.155202} {\bibfield  {journal} {\bibinfo
  {journal} {Phys. Rev. B}\ }\textbf {\bibinfo {volume} {98}},\ \bibinfo
  {pages} {155202} (\bibinfo {year} {2018})}\BibitemShut {NoStop}%
\bibitem [{\citenamefont {Yang}\ \emph {et~al.}(2020)\citenamefont {Yang},
  \citenamefont {Wang}, \citenamefont {Tao}, \citenamefont {Yang},
  \citenamefont {Zhang}, \citenamefont {Ai},\ and\ \citenamefont
  {Deng}}]{yang20}%
  \BibitemOpen
  \bibfield  {author} {\bibinfo {author} {\bibfnamefont {Z.-S.}\ \bibnamefont
  {Yang}}, \bibinfo {author} {\bibfnamefont {Y.-X.}\ \bibnamefont {Wang}},
  \bibinfo {author} {\bibfnamefont {M.-J.}\ \bibnamefont {Tao}}, \bibinfo
  {author} {\bibfnamefont {W.}~\bibnamefont {Yang}}, \bibinfo {author}
  {\bibfnamefont {M.}~\bibnamefont {Zhang}}, \bibinfo {author} {\bibfnamefont
  {Q.}~\bibnamefont {Ai}},\ and\ \bibinfo {author} {\bibfnamefont {F.-G.}\
  \bibnamefont {Deng}},\ }\bibfield  {title} {\bibinfo {title} {Longitudinal
  relaxation of a nitrogen-vacancy center in a spin bath by generalized
  cluster-correlation expansion method},\ }\href
  {https://doi.org/https://doi.org/10.1016/j.aop.2019.168063} {\bibfield
  {journal} {\bibinfo  {journal} {Annals of Physics}\ }\textbf {\bibinfo
  {volume} {413}},\ \bibinfo {pages} {168063} (\bibinfo {year}
  {2020})}\BibitemShut {NoStop}%
\bibitem [{\citenamefont {Onizhuk}\ \emph {et~al.}(2021)\citenamefont
  {Onizhuk}, \citenamefont {Miao}, \citenamefont {Blanton}, \citenamefont {Ma},
  \citenamefont {Anderson}, \citenamefont {Bourassa}, \citenamefont
  {Awschalom},\ and\ \citenamefont {Galli}}]{onizhuk21}%
  \BibitemOpen
  \bibfield  {author} {\bibinfo {author} {\bibfnamefont {M.}~\bibnamefont
  {Onizhuk}}, \bibinfo {author} {\bibfnamefont {K.~C.}\ \bibnamefont {Miao}},
  \bibinfo {author} {\bibfnamefont {J.~P.}\ \bibnamefont {Blanton}}, \bibinfo
  {author} {\bibfnamefont {H.}~\bibnamefont {Ma}}, \bibinfo {author}
  {\bibfnamefont {C.~P.}\ \bibnamefont {Anderson}}, \bibinfo {author}
  {\bibfnamefont {A.}~\bibnamefont {Bourassa}}, \bibinfo {author}
  {\bibfnamefont {D.~D.}\ \bibnamefont {Awschalom}},\ and\ \bibinfo {author}
  {\bibfnamefont {G.}~\bibnamefont {Galli}},\ }\bibfield  {title} {\bibinfo
  {title} {Probing the coherence of solid-state qubits at avoided crossings},\
  }\href {https://doi.org/10.1103/PRXQuantum.2.010311} {\bibfield  {journal}
  {\bibinfo  {journal} {PRX Quantum}\ }\textbf {\bibinfo {volume} {2}},\
  \bibinfo {pages} {010311} (\bibinfo {year} {2021})}\BibitemShut {NoStop}%
\bibitem [{\citenamefont {Ghosh}\ \emph {et~al.}(2021)\citenamefont {Ghosh},
  \citenamefont {Ma}, \citenamefont {Onizhuk}, \citenamefont {Gavini},\ and\
  \citenamefont {Galli}}]{ghosh21}%
  \BibitemOpen
  \bibfield  {author} {\bibinfo {author} {\bibfnamefont {K.}~\bibnamefont
  {Ghosh}}, \bibinfo {author} {\bibfnamefont {H.}~\bibnamefont {Ma}}, \bibinfo
  {author} {\bibfnamefont {M.}~\bibnamefont {Onizhuk}}, \bibinfo {author}
  {\bibfnamefont {V.}~\bibnamefont {Gavini}},\ and\ \bibinfo {author}
  {\bibfnamefont {G.}~\bibnamefont {Galli}},\ }\bibfield  {title} {\bibinfo
  {title} {Spin–spin interactions in defects in solids from mixed
  all-electron and pseudopotential first-principles calculations},\ }\href
  {https://doi.org/10.1038/s41524-021-00590-w} {\bibfield  {journal} {\bibinfo
  {journal} {npj Computational Materials}\ }\textbf {\bibinfo {volume} {7}},\
  \bibinfo {pages} {123} (\bibinfo {year} {2021})}\BibitemShut {NoStop}%
	\bibitem [{\citenamefont {addison}(2017)}]{addison02}%
	    \BibitemOpen
	    \bibfield  {author} {\bibinfo {author} {\bibfnamefont {P.~S.}\ \bibnamefont
	    {Addison}},\ }\href {https://doi.org/10.1201/9781420033397} {\emph
	    {\bibinfo {title} {The illustrated wavelet transform handbook: introductory theory and applications in science, engineering, medicine and finance}}}\ (\bibinfo
	    {publisher} {CRC press}, \ \bibinfo {address} {Boca Raton}, \ \bibinfo {year}
	    {2017})\BibitemShut {NoStop}%
\bibitem [{\citenamefont {Phillies}\ and\ \citenamefont
  {Stott}(1995{\natexlab{a}})}]{phillies95-1}%
  \BibitemOpen
  \bibfield  {author} {\bibinfo {author} {\bibfnamefont {G.~D.~J.}\
  \bibnamefont {Phillies}}\ and\ \bibinfo {author} {\bibfnamefont
  {J.}~\bibnamefont {Stott}},\ }\bibfield  {title} {\bibinfo {title} {Wavelet
  analysis of ising model spin dynamics},\ }\href
  {https://doi.org/10.1063/1.168539} {\bibfield  {journal} {\bibinfo  {journal}
  {Computers in Physics}\ }\textbf {\bibinfo {volume} {9}},\ \bibinfo {pages}
  {97} (\bibinfo {year} {1995}{\natexlab{a}})},\ \Eprint
  {https://arxiv.org/abs/https://aip.scitation.org/doi/pdf/10.1063/1.168539}
  {https://aip.scitation.org/doi/pdf/10.1063/1.168539} \BibitemShut {NoStop}%
\bibitem [{\citenamefont {Phillies}\ and\ \citenamefont
  {Stott}(1995{\natexlab{b}})}]{phillies95-2}%
  \BibitemOpen
  \bibfield  {author} {\bibinfo {author} {\bibfnamefont {G.~D.~J.}\
  \bibnamefont {Phillies}}\ and\ \bibinfo {author} {\bibfnamefont
  {J.}~\bibnamefont {Stott}},\ }\bibfield  {title} {\bibinfo {title}
  {Mori–zwanzig–daubechies decomposition of ising‐model monte carlo
  dynamics},\ }\href {https://doi.org/10.1063/1.168527} {\bibfield  {journal}
  {\bibinfo  {journal} {Computers in Physics}\ }\textbf {\bibinfo {volume}
  {9}},\ \bibinfo {pages} {225} (\bibinfo {year} {1995}{\natexlab{b}})},\
  \Eprint
  {https://arxiv.org/abs/https://aip.scitation.org/doi/pdf/10.1063/1.168527}
  {https://aip.scitation.org/doi/pdf/10.1063/1.168527} \BibitemShut {NoStop}%
\bibitem [{\citenamefont {Phillies}(1996)}]{phillies96}%
  \BibitemOpen
  \bibfield  {author} {\bibinfo {author} {\bibfnamefont {G.~D.~J.}\
  \bibnamefont {Phillies}},\ }\bibfield  {title} {\bibinfo {title} {Wavelets: a
  new alternative to fourier transforms},\ }\href
  {https://doi.org/10.1063/1.168573} {\bibfield  {journal} {\bibinfo  {journal}
  {Computers in Physics}\ }\textbf {\bibinfo {volume} {10}},\ \bibinfo {pages}
  {247} (\bibinfo {year} {1996})},\ \Eprint
  {https://arxiv.org/abs/https://aip.scitation.org/doi/pdf/10.1063/1.168573}
  {https://aip.scitation.org/doi/pdf/10.1063/1.168573} \BibitemShut {NoStop}%
\bibitem [{\citenamefont {Abbott}(2016)}]{abbot16}%
  \BibitemOpen
  \bibfield  {author} {\bibinfo {author} {\bibfnamefont {B.~P.~t.}\
  \bibnamefont {Abbott}} (\bibinfo {collaboration} {LIGO Scientific
  Collaboration and Virgo Collaboration}),\ }\bibfield  {title} {\bibinfo
  {title} {Observation of gravitational waves from a binary black hole
  merger},\ }\href {https://doi.org/10.1103/PhysRevLett.116.061102} {\bibfield
  {journal} {\bibinfo  {journal} {Phys. Rev. Lett.}\ }\textbf {\bibinfo
  {volume} {116}},\ \bibinfo {pages} {061102} (\bibinfo {year}
  {2016})}\BibitemShut {NoStop}%
\bibitem [{\citenamefont {Pham}\ and\ \citenamefont {Meignen}(2017)}]{pham17}%
  \BibitemOpen
  \bibfield  {author} {\bibinfo {author} {\bibfnamefont {D.-H.}\ \bibnamefont
  {Pham}}\ and\ \bibinfo {author} {\bibfnamefont {S.}~\bibnamefont {Meignen}},\
  }\bibfield  {title} {\bibinfo {title} {High-order synchrosqueezing transform
  for multicomponent signals analysis—with an application to
  gravitational-wave signal},\ }\href
  {https://doi.org/10.1109/TSP.2017.2686355} {\bibfield  {journal} {\bibinfo
  {journal} {IEEE Transactions on Signal Processing}\ }\textbf {\bibinfo
  {volume} {65}},\ \bibinfo {pages} {3168} (\bibinfo {year}
  {2017})}\BibitemShut {NoStop}%
\bibitem [{\citenamefont {Daubechies}\ \emph {et~al.}(2011)\citenamefont
  {Daubechies}, \citenamefont {L.},\ and\ \citenamefont {W.}}]{daubechies11}%
  \BibitemOpen
  \bibfield  {author} {\bibinfo {author} {\bibfnamefont {I.}~\bibnamefont
  {Daubechies}}, \bibinfo {author} {\bibfnamefont {J.}~\bibnamefont {L.}},\
  and\ \bibinfo {author} {\bibfnamefont {H.-T.}\ \bibnamefont {W.}},\
  }\bibfield  {title} {\bibinfo {title} {Synchrosqueezed wavelet transforms: An
  empirical mode decomposition-like tool},\ }\href
  {https://doi.org/https://doi.org/10.1016/j.acha.2010.08.002} {\bibfield
  {journal} {\bibinfo  {journal} {Applied and Computational Harmonic Analysis}\
  }\textbf {\bibinfo {volume} {30}},\ \bibinfo {pages} {243} (\bibinfo {year}
  {2011})}\BibitemShut {NoStop}%
\bibitem [{\citenamefont {Tary}\ \emph {et~al.}(2018)\citenamefont {Tary},
  \citenamefont {Herrera},\ and\ \citenamefont {van~der Baan}}]{tary18}%
  \BibitemOpen
  \bibfield  {author} {\bibinfo {author} {\bibfnamefont {J.~B.}\ \bibnamefont
  {Tary}}, \bibinfo {author} {\bibfnamefont {R.~H.}\ \bibnamefont {Herrera}},\
  and\ \bibinfo {author} {\bibfnamefont {M.}~\bibnamefont {van~der Baan}},\
  }\bibfield  {title} {\bibinfo {title} {Analysis of time-varying signals using
  continuous wavelet and synchrosqueezed transforms},\ }\href
  {https://doi.org/10.1098/rsta.2017.0254} {\bibfield  {journal} {\bibinfo
  {journal} {Philosophical Transactions of the Royal Society A: Mathematical,
  Physical and Engineering Sciences}\ }\textbf {\bibinfo {volume} {376}},\
  \bibinfo {pages} {20170254} (\bibinfo {year} {2018})},\ \Eprint
  {https://arxiv.org/abs/https://royalsocietypublishing.org/doi/pdf/10.1098/rsta.2017.0254}
  {https://royalsocietypublishing.org/doi/pdf/10.1098/rsta.2017.0254}
  \BibitemShut {NoStop}%
\bibitem [{\citenamefont {Witzel}\ \emph {et~al.}(2014)\citenamefont {Witzel},
  \citenamefont {Young},\ and\ \citenamefont {Das~Sarma}}]{witzel14}%
  \BibitemOpen
  \bibfield  {author} {\bibinfo {author} {\bibfnamefont {W.~M.}\ \bibnamefont
  {Witzel}}, \bibinfo {author} {\bibfnamefont {K.}~\bibnamefont {Young}},\ and\
  \bibinfo {author} {\bibfnamefont {S.}~\bibnamefont {Das~Sarma}},\ }\bibfield
  {title} {\bibinfo {title} {Converting a real quantum spin bath to an
  effective classical noise acting on a central spin},\ }\href
  {https://doi.org/10.1103/PhysRevB.90.115431} {\bibfield  {journal} {\bibinfo
  {journal} {Phys. Rev. B}\ }\textbf {\bibinfo {volume} {90}},\ \bibinfo
  {pages} {115431} (\bibinfo {year} {2014})}\BibitemShut {NoStop}%
\bibitem [{\citenamefont {Ma}\ \emph {et~al.}()\citenamefont {Ma},
  \citenamefont {Wolfowicz}, \citenamefont {Li}, \citenamefont {Morton},\ and\
  \citenamefont {Liu}}]{ma15}%
  \BibitemOpen
  \bibfield  {author} {\bibinfo {author} {\bibfnamefont {W.-L.}\ \bibnamefont
  {Ma}}, \bibinfo {author} {\bibfnamefont {G.}~\bibnamefont {Wolfowicz}},
  \bibinfo {author} {\bibfnamefont {S.-S.}\ \bibnamefont {Li}}, \bibinfo
  {author} {\bibfnamefont {J.~J.~L.}\ \bibnamefont {Morton}},\ and\ \bibinfo
  {author} {\bibfnamefont {R.-B.}\ \bibnamefont {Liu}},\ }\bibfield  {title}
  {\bibinfo {title} {Classical nature of nuclear spin noise near clock
  transitions of bi donors in silicon},\ }\href {https://link.aps.org/doi/10.1103/PhysRevB.92.161403} {\bibfield  {journal}
  {\bibinfo  {journal} {Phys. Rev. B}\ }\textbf {\bibinfo {volume} {92}},\
  \bibinfo {pages} {161403} (\bibinfo {year} {2015})}\BibitemShut {NoStop}%
\bibitem [{\citenamefont {Witzel}\ and\ \citenamefont
  {Das~Sarma}(2008)}]{witzel08}%
  \BibitemOpen
  \bibfield  {author} {\bibinfo {author} {\bibfnamefont {W.~M.}\ \bibnamefont
  {Witzel}}\ and\ \bibinfo {author} {\bibfnamefont {S.}~\bibnamefont
  {Das~Sarma}},\ }\bibfield  {title} {\bibinfo {title} {Wavefunction
  considerations for the central spin decoherence problem in a nuclear spin
  bath},\ }\href {https://doi.org/10.1103/PhysRevB.77.165319} {\bibfield
  {journal} {\bibinfo  {journal} {Phys. Rev. B}\ }\textbf {\bibinfo {volume}
  {77}},\ \bibinfo {pages} {165319} (\bibinfo {year} {2008})}\BibitemShut
  {NoStop}%
\bibitem [{\citenamefont {Giedke}\ \emph {et~al.}(2006)\citenamefont {Giedke},
  \citenamefont {Taylor}, \citenamefont {D'Alessandro}, \citenamefont {Lukin},\
  and\ \citenamefont {Imamo\ifmmode~\breve{g}\else \u{g}\fi{}lu}}]{giedke06}%
  \BibitemOpen
  \bibfield  {author} {\bibinfo {author} {\bibfnamefont {G.}~\bibnamefont
  {Giedke}}, \bibinfo {author} {\bibfnamefont {J.~M.}\ \bibnamefont {Taylor}},
  \bibinfo {author} {\bibfnamefont {D.}~\bibnamefont {D'Alessandro}}, \bibinfo
  {author} {\bibfnamefont {M.~D.}\ \bibnamefont {Lukin}},\ and\ \bibinfo
  {author} {\bibfnamefont {A.}~\bibnamefont {Imamo\ifmmode~\breve{g}\else
  \u{g}\fi{}lu}},\ }\bibfield  {title} {\bibinfo {title} {Quantum measurement
  of a mesoscopic spin ensemble},\ }\href
  {https://doi.org/10.1103/PhysRevA.74.032316} {\bibfield  {journal} {\bibinfo
  {journal} {Phys. Rev. A}\ }\textbf {\bibinfo {volume} {74}},\ \bibinfo
  {pages} {032316} (\bibinfo {year} {2006})}\BibitemShut {NoStop}%
\bibitem [{\citenamefont {Slichter}(1990)}]{slichter90}%
  \BibitemOpen
  \bibfield  {author} {\bibinfo {author} {\bibfnamefont {C.~P.}\ \bibnamefont
  {Slichter}},\ }\href {https://doi.org/10.1007/978-3-662-09441-9} {\emph
  {\bibinfo {title} {Principles of Magnetic Resonance}}}\ (\bibinfo
  {publisher} {Springer Berlin Heidelberg},\ \bibinfo {year}
  {1990})\BibitemShut {NoStop}%
\bibitem [{\citenamefont {Abragam}(1961)}]{abragam99}%
	    \BibitemOpen
	    \bibfield  {author} {\bibinfo {author} {\bibfnamefont {A.}~\bibnamefont
	    {Abragam}},\ }\href@noop {} {\emph {\bibinfo {title} {The principles of
	    nuclear magnetism}}}\ (\bibinfo  {publisher} {Oxford University Press}, \ \bibinfo {address} {Oxford}, \ \bibinfo
	    {year} {1961})\BibitemShut {NoStop}%
\bibitem [{\citenamefont {Schliemann}\ \emph {et~al.}(2003)\citenamefont
  {Schliemann}, \citenamefont {Khaetskii},\ and\ \citenamefont
  {Loss}}]{schliemann03}%
  \BibitemOpen
  \bibfield  {author} {\bibinfo {author} {\bibfnamefont {J.}~\bibnamefont
  {Schliemann}}, \bibinfo {author} {\bibfnamefont {A.}~\bibnamefont
  {Khaetskii}},\ and\ \bibinfo {author} {\bibfnamefont {D.}~\bibnamefont
  {Loss}},\ }\bibfield  {title} {\bibinfo {title} {Electron spin dynamics in
  quantum dots and related nanostructures due to hyperfine interaction with
  nuclei},\ }\href {http://stacks.iop.org/0953-8984/15/i=50/a=R01} {\bibfield
  {journal} {\bibinfo  {journal} {Journal of Physics: Condensed Matter}\
  }\textbf {\bibinfo {volume} {15}},\ \bibinfo {pages} {R1809} (\bibinfo {year}
  {2003})}\BibitemShut {NoStop}%
\bibitem [{\citenamefont {G\"uldeste}\ and\ \citenamefont
  {Bulutay}(2018)}]{Guldeste18}%
  \BibitemOpen
  \bibfield  {author} {\bibinfo {author} {\bibfnamefont {E.~T.}\ \bibnamefont
  {G\"uldeste}}\ and\ \bibinfo {author} {\bibfnamefont {C.}~\bibnamefont
  {Bulutay}},\ }\bibfield  {title} {\bibinfo {title} {Loschmidt echo driven by
  hyperfine and electric-quadrupole interactions in nanoscale nuclear spin
  baths},\ }\href {https://doi.org/10.1103/PhysRevB.98.085202} {\bibfield
  {journal} {\bibinfo  {journal} {Phys. Rev. B}\ }\textbf {\bibinfo {volume}
  {98}},\ \bibinfo {pages} {085202} (\bibinfo {year} {2018})}\BibitemShut
  {NoStop}%
\bibitem [{\citenamefont {Giri}\ \emph {et~al.}(2019)\citenamefont {Giri},
  \citenamefont {Dorigoni}, \citenamefont {Tambalo}, \citenamefont {Gorrini},\
  and\ \citenamefont {Bifone}}]{giri19}%
  \BibitemOpen
  \bibfield  {author} {\bibinfo {author} {\bibfnamefont {R.}~\bibnamefont
  {Giri}}, \bibinfo {author} {\bibfnamefont {C.}~\bibnamefont {Dorigoni}},
  \bibinfo {author} {\bibfnamefont {S.}~\bibnamefont {Tambalo}}, \bibinfo
  {author} {\bibfnamefont {F.}~\bibnamefont {Gorrini}},\ and\ \bibinfo {author}
  {\bibfnamefont {A.}~\bibnamefont {Bifone}},\ }\bibfield  {title} {\bibinfo
  {title} {Selective measurement of charge dynamics in an ensemble of
  nitrogen-vacancy centers in nanodiamond and bulk diamond},\ }\href
  {https://doi.org/10.1103/PhysRevB.99.155426} {\bibfield  {journal} {\bibinfo
  {journal} {Phys. Rev. B}\ }\textbf {\bibinfo {volume} {99}},\ \bibinfo
  {pages} {155426} (\bibinfo {year} {2019})}\BibitemShut {NoStop}%
\bibitem [{\citenamefont {Mallat}(2009)}]{stephane09}%
  \BibitemOpen
  \bibfield  {author} {\bibinfo {author} {\bibfnamefont {S.}~\bibnamefont
  {Mallat}},\ }\bibfield  {title} {\bibinfo {title} {Sparse representations},\
  }in\ \href
  {https://doi.org/https://doi.org/10.1016/B978-0-12-374370-1.00005-7} {\emph
  {\bibinfo {booktitle} {A Wavelet Tour of Signal Processing (Third
  Edition)}}},\ \bibinfo {editor} {edited by\ \bibinfo {editor} {\bibfnamefont
  {S.}~\bibnamefont {Mallat}}}\ (\bibinfo  {publisher} {Academic Press},\
  \bibinfo {address} {Boston},\ \bibinfo {year} {2009})\ \bibinfo {edition}
  {third edition}\ ed.,\ pp.\ \bibinfo {pages} {1--31}\BibitemShut {NoStop}%
\bibitem [{\citenamefont {Meignen}\ \emph {et~al.}(2012)\citenamefont
  {Meignen}, \citenamefont {Oberlin},\ and\ \citenamefont
  {McLaughlin}}]{sylvain12}%
  \BibitemOpen
  \bibfield  {author} {\bibinfo {author} {\bibfnamefont {S.}~\bibnamefont
  {Meignen}}, \bibinfo {author} {\bibfnamefont {T.}~\bibnamefont {Oberlin}},\
  and\ \bibinfo {author} {\bibfnamefont {S.}~\bibnamefont {McLaughlin}},\
  }\bibfield  {title} {\bibinfo {title} {A new algorithm for multicomponent
  signals analysis based on synchrosqueezing: With an application to signal
  sampling and denoising},\ }\href {https://doi.org/10.1109/TSP.2012.2212891}
  {\bibfield  {journal} {\bibinfo  {journal} {IEEE Transactions on Signal
  Processing}\ }\textbf {\bibinfo {volume} {60}},\ \bibinfo {pages} {5787}
  (\bibinfo {year} {2012})}\BibitemShut {NoStop}%
\bibitem [{\citenamefont {Witzel}\ \emph {et~al.}(2012)\citenamefont {Witzel},
  \citenamefont {Carroll}, \citenamefont {Cywi\ifmmode~\acute{n}\else
  \'{n}\fi{}ski},\ and\ \citenamefont {Das~Sarma}}]{witzel12}%
  \BibitemOpen
  \bibfield  {author} {\bibinfo {author} {\bibfnamefont {W.~M.}\ \bibnamefont
  {Witzel}}, \bibinfo {author} {\bibfnamefont {M.~S.}\ \bibnamefont {Carroll}},
  \bibinfo {author} {\bibfnamefont {L.}~\bibnamefont
  {Cywi\ifmmode~\acute{n}\else \'{n}\fi{}ski}},\ and\ \bibinfo {author}
  {\bibfnamefont {S.}~\bibnamefont {Das~Sarma}},\ }\bibfield  {title} {\bibinfo
  {title} {Quantum decoherence of the central spin in a sparse system of
  dipolar coupled spins},\ }\href {https://doi.org/10.1103/PhysRevB.86.035452}
  {\bibfield  {journal} {\bibinfo  {journal} {Phys. Rev. B}\ }\textbf {\bibinfo
  {volume} {86}},\ \bibinfo {pages} {035452} (\bibinfo {year}
  {2012})}\BibitemShut {NoStop}%
\bibitem [{\citenamefont {Bulutay}(2012)}]{bulutay12}%
  \BibitemOpen
  \bibfield  {author} {\bibinfo {author} {\bibfnamefont {C.}~\bibnamefont
  {Bulutay}},\ }\bibfield  {title} {\bibinfo {title} {Quadrupolar spectra of
  nuclear spins in strained in${}_{x}$ga${}_{1\ensuremath{-}x}$as quantum
  dots},\ }\href {https://doi.org/10.1103/PhysRevB.85.115313} {\bibfield
  {journal} {\bibinfo  {journal} {Phys. Rev. B}\ }\textbf {\bibinfo {volume}
  {85}},\ \bibinfo {pages} {115313} (\bibinfo {year} {2012})}\BibitemShut
  {NoStop}%
\bibitem [{\citenamefont {Sza{\'{n}}kowski}\ \emph {et~al.}(2017)\citenamefont
  {Sza{\'{n}}kowski}, \citenamefont {Ramon}, \citenamefont {Krzywda},
  \citenamefont {Kwiatkowski},\ and\ \citenamefont
  {Cywi{\'{n}}ski}}]{szankowski17}%
  \BibitemOpen
  \bibfield  {author} {\bibinfo {author} {\bibfnamefont {P.}~\bibnamefont
  {Sza{\'{n}}kowski}}, \bibinfo {author} {\bibfnamefont {G.}~\bibnamefont
  {Ramon}}, \bibinfo {author} {\bibfnamefont {J.}~\bibnamefont {Krzywda}},
  \bibinfo {author} {\bibfnamefont {D.}~\bibnamefont {Kwiatkowski}},\ and\
  \bibinfo {author} {\bibfnamefont {{\L}.}~\bibnamefont {Cywi{\'{n}}ski}},\
  }\bibfield  {title} {\bibinfo {title} {Environmental noise spectroscopy with
  qubits subjected to dynamical decoupling},\ }\href
  {https://doi.org/10.1088/1361-648x/aa7648} {\bibfield  {journal} {\bibinfo
  {journal} {Journal of Physics: Condensed Matter}\ }\textbf {\bibinfo {volume}
  {29}},\ \bibinfo {pages} {333001} (\bibinfo {year} {2017})}\BibitemShut
  {NoStop}%
\bibitem [{\citenamefont {Ramon}(2015)}]{ramon15}%
  \BibitemOpen
  \bibfield  {author} {\bibinfo {author} {\bibfnamefont {G.}~\bibnamefont
  {Ramon}},\ }\bibfield  {title} {\bibinfo {title} {Non-gaussian signatures and
  collective effects in charge noise affecting a dynamically decoupled qubit},\
  }\href {https://doi.org/10.1103/PhysRevB.92.155422} {\bibfield  {journal}
  {\bibinfo  {journal} {Phys. Rev. B}\ }\textbf {\bibinfo {volume} {92}},\
  \bibinfo {pages} {155422} (\bibinfo {year} {2015})}\BibitemShut {NoStop}%
\bibitem [{\citenamefont {Nason}\ and\ \citenamefont
  {Silverman}(1995)}]{nason95}%
  \BibitemOpen
  \bibfield  {author} {\bibinfo {author} {\bibfnamefont {G.~P.}\ \bibnamefont
  {Nason}}\ and\ \bibinfo {author} {\bibfnamefont {B.~W.}\ \bibnamefont
  {Silverman}},\ }\bibinfo {title} {The stationary wavelet transform and some
  statistical applications},\ in\ \href
  {https://doi.org/10.1007/978-1-4612-2544-7_17} {\emph {\bibinfo {booktitle}
  {Wavelets and Statistics}}},\ \bibinfo {editor} {edited by\ \bibinfo {editor}
  {\bibfnamefont {A.}~\bibnamefont {Antoniadis}}\ and\ \bibinfo {editor}
  {\bibfnamefont {G.}~\bibnamefont {Oppenheim}}}\ (\bibinfo  {publisher}
  {Springer New York},\ \bibinfo {address} {New York, NY},\ \bibinfo {year}
  {1995})\ pp.\ \bibinfo {pages} {281--299}\BibitemShut {NoStop}%
\bibitem [{\citenamefont {Gangloff}\ \emph {et~al.}(2021)\citenamefont
  {Gangloff}, \citenamefont {Zaporski}, \citenamefont {Bodey}, \citenamefont
  {Bachorz}, \citenamefont {Jackson}, \citenamefont {{\'E}thier-Majcher},
  \citenamefont {Lang}, \citenamefont {Clarke}, \citenamefont {Hugues},
  \citenamefont {Le~Gall} \emph {et~al.}}]{gangloff21}%
  \BibitemOpen
  \bibfield  {author} {\bibinfo {author} {\bibfnamefont {D.~A.}\ \bibnamefont
  {Gangloff}}, \bibinfo {author} {\bibfnamefont {L.}~\bibnamefont {Zaporski}},
  \bibinfo {author} {\bibfnamefont {J.~H.}\ \bibnamefont {Bodey}}, \bibinfo
  {author} {\bibfnamefont {C.}~\bibnamefont {Bachorz}}, \bibinfo {author}
  {\bibfnamefont {D.~M.}\ \bibnamefont {Jackson}}, \bibinfo {author}
  {\bibfnamefont {G.}~\bibnamefont {{\'E}thier-Majcher}}, \bibinfo {author}
  {\bibfnamefont {C.}~\bibnamefont {Lang}}, \bibinfo {author} {\bibfnamefont
  {E.}~\bibnamefont {Clarke}}, \bibinfo {author} {\bibfnamefont
  {M.}~\bibnamefont {Hugues}}, \bibinfo {author} {\bibfnamefont
  {C.}~\bibnamefont {Le~Gall}}, \emph {et~al.},\ }\bibfield  {title} {\bibinfo
  {title} {Witnessing quantum correlations in a nuclear ensemble via an
  electron spin qubit},\ }\href {https://doi.org/s41567-021-01344-7} {\bibfield
   {journal} {\bibinfo  {journal} {Nature Physics}\ ,\ \bibinfo {pages} {1}}
  (\bibinfo {year} {2021})}\BibitemShut {NoStop}%
\end{thebibliography}
\end{document}